\documentclass[twocolumn]{aastex63}

\received{June 1, 2019}
\revised{January 10, 2019}
\accepted{\today}
\shorttitle{}
\shortauthors{}

\graphicspath{{./}{figures/}}

\usepackage{fdsymbol}
\usepackage{soul}
\usepackage{dcolumn}
\usepackage{booktabs}
\usepackage{tabularx}
\usepackage{rotating}
\usepackage{gensymb}
\usepackage{placeins}
\usepackage{etoolbox}
\usepackage{xcolor}
\newcolumntype{Y}{>{\centering\arraybackslash}X}

\begin{document}

\title{Discovery and Timing of 49 Pulsars from the Arecibo 327-MHz Drift Survey}

\correspondingauthor{Timothy E. E. Olszanski}
\email{teo0008@mix.wvu.edu}
\author{Timothy E. E. Olszanski}
\affiliation{Department of Physics and Astronomy, West Virginia University, P.O. Box 6315, Morgantown, WV 26506, USA}
\affiliation{Center for Gravitational Waves and Cosmology, West Virginia University, Chestnut Ridge Research Building, Morgantown, WV 26506, USA}
\author{Evan F. Lewis}
\affiliation{Department of Physics and Astronomy, West Virginia University, P.O. Box 6315, Morgantown, WV 26506, USA}
\affiliation{Center for Gravitational Waves and Cosmology, West Virginia University, Chestnut Ridge Research Building, Morgantown, WV 26506, USA}
\author{Julia S. Deneva}
\affiliation{George Mason University, Fairfax, VA 22030, USA}
\author{Maura A. McLaughlin}
\affiliation{Department of Physics and Astronomy, West Virginia University, P.O. Box 6315, Morgantown, WV 26506, USA}
\affiliation{Center for Gravitational Waves and Cosmology, West Virginia University, Chestnut Ridge Research Building, Morgantown, WV 26506, USA}
\author{Kevin Stovall}
\author{Paulo C. C. Freire}
\affiliation{Max-Planck-Institut f\"ur Radioastronomie, Bonn, Germany}
\author{Benetge B. P. Perera}
\affiliation{Florida Space Institute, University of Central Florida, 12354 Research Parkway, Orlando, FL 32826, USA}
\author{Manjari Bagchi}
\affiliation{The Institute of Mathematical Sciences, Chennai, India 600113}
\affiliation{Homi Bhabha National Institute, Mumbai, India 400094}
\author{Jose G. Martinez}
\affiliation{Max-Planck-Institut f\"ur Radioastronomie, Bonn, Germany}

\begin{abstract}
We present 18 pulsar discoveries from the AO327 pulsar survey, along with their timing solutions and those for an additional 31  AO327-discovered pulsars. Timing solutions were constructed using observations from a follow-up timing campaign taken between the periods of 2013 --- 2019 using the Arecibo Observatory's 327-MHz receiver. Aside from PSR J0916+0658, an isolated pulsar that shows evidence for partial recycling, the remaining discoveries are non-recycled pulsars. We present a brief census of emission features for all pulsars with the following standouts. PSR~J1942+0142 is found to exhibit the very rare phenomenon of subpulse bi-drifting and PSR~J0225+1727 has an interpulse. We also report distance estimates using the NE2001, YMW16, and NE2025 Galactic electron density models, and identify at least 10 sources where either one or more models underestimate the maximum Galactic line of sight dispersion measure. We compare our discoveries with those of the GBNCC survey, finding that off the Galactic plane, the majority of failures arise from YMW16, while in the Galactic plane, NE2025 shows a marginal degradation of performance relative to NE2001.
\end{abstract}


\section{Introduction}

Since their discovery by Jocelyn Bell over 50 years ago \citep{1968Natur.217..709H}, radio pulsars have proven themselves to be immensely useful astrophysical laboratories, ranging from the nature of their multi-wavelength emission and corresponding fundamental plasma physics \citep{Harding_2017}, pulsar and neutron star populations and their consequences for stellar evolution \citep{https://doi.org/10.48550/arxiv.1903.06526}, stellar evolution in binaries \citep{Tauris2017,Tauris2023}, and as probes of the interstellar medium \citep{Turner_2021}. They have provided a number of fundamental physics advancements, the most prominent of which are evidence for a nanohertz gravitational wave background via the use of pulsar timing arrays \citep[PTAs,][]{Agazie_2023},  tests of general relativity (GR) and fundamental properties of gravity (e.g., \citealt{Kramer2021}, for a general review see \citealt{2024LRR....27....5F}), and tests of the properties of the strong nuclear force via their equation of state \citep[EOS,][]{OzelFreire2016,Lattimer2021}. Most of this work is done in the radio band, in which coherent, pulsed emission is produced from charged particle acceleration along the magnetic axis.

Although more than 4,300 pulsars are known\footnote{\url{https://www.atnf.csiro.au/research/pulsar/psrcat/}} \citep{mht+05}, extensive radio pulsar surveys carried out world-wide \citep[e.g.,][]{mlc+01,cfl+06,kjs+10,slr+14,hww+21,rgf+21} continue to produce new discoveries. In the last decade alone, these include ultra long period pulsars \citep{2022NatAs...6..828C,PhysRevD.107.L081301},  high-mass pulsars \citep{2020NatAs...4...72C,Fonseca2021}, highly relativistic binary systems \citep{2023Natur.620..961P} and possibly the first pulsar - black hole system \citep{Barr_2024}. The Five-hundred-meter Aperture Spherical Telescope (FAST) in China has discovered 751 new pulsars in only a few years of operation\footnote{\url{http://zmtt.bao.ac.cn/GPPS/GPPSnewPSR.html}}\citep{Han_2025}.
Another major impetus for ongoing pulsar surveys is to find suitable sources for inclusion in PTAs \citep{Alam_2020,2019MNRAS.490.4666P,ccg+21,krh+20}. With the recent evidence for nanohertz gravitational waves presented by the NANOGrav collaboration \citep{Agazie_2023} and other PTAs \citep{all_pta,epta_paper,Reardon_2023}, pulsar surveys serve an ever more valuable role to PTAs by identifying pulsars ideal for precision timing.

The radio sky is not limited only to pulsars. Surveys are sensitive to transient sources such as fast radio bursts \citep[FRBs,][]{annurev:/content/journals/10.1146/annurev-astro-091918-104501}, which originate from outside the Galaxy, rotating radio transients \citep[RRATs,][]{rrat1, keane2011rotatingradiotransients,rrat_thornton} -- neutron stars that emit sporadic radio pulses, and ultra long period transients \citep{2024NatAs...8.1159C}. New survey discoveries help to improve the census of neutron stars as well as the modeling of the Galaxy's free electron distribution. At low frequencies, dispersion and scattering become more pronounced, limiting sensitivity to only a fraction of the population. As most pulsar spectra peak below 500 MHz, low-frequency surveys are especially sensitive to both nearby and distant pulsars that lie on lines of sight with low electron densities. These lines of sight are very useful for probing the limitations of current electron density models, and are in turn useful for refining distance estimates for FRBs. With its unmatched sensitivity to the faintest sources, Arecibo was an ideal telescope for this work.

Pulsars divide into roughly two observational classes; those with and without a history of accretion. Pulsars where significant accretion has taken place have shorter spin periods as compared to younger, non-recycled pulsars. Millisecond pulsar (MSP) pulse profiles are far more complex than non-recycled pulsars. It is likely accretion modifies the structure of MSP magnetospheres, and by extension, emission regions; there is also strong evidence that MSP radio emission partly arises from the light cylinder rather than the polar cap \citep{kramer2025radioemissionlightcylinder}. There are few MSPs from which single pulses are detectable. In contrast, non-recycled pulsars, in particular older pulsars with longer spin periods, show strong evidence for a more organized emission geometry \citep{core_cone}, and single pulses are more easily detectable. This single-pulse radio emission is characterized by a rich collection of behaviors that in turn reflect the changing physical conditions of the underlying fundamental magnetospheric plasma physics. Broadly, these behaviors can be subdivided into two categories; those that involve amplitude modulation of some form e.g. nulling, giant pulses, mode changing, periodic amplitude modulation etc, \citep{nulling,giant_pulses,mode_changing,Basu_2025}, and those that relate to apparent motion of emission across the beam such as subpulse drift \citep{subpulse_drift}. Studies of these behaviors serve as useful observational constraints for theorists.
%Young pulsars also exhibit some of these behaviors, but not to the same degree. With regards to MSPs, it is an open question how similar they are to slow pulsars. While in a few isolated cases, MSPs have been seen to show similar single-pulse behaviors \citep{Mahajan_2018,msp_ben} and emission geometries \citep{Rankin_2017}, most MSPs have complex emission geometries that are significantly more complex than slow pulsars and also have very few detectable single pulses using even the most sensitive instruments which limits the degree to which the two groups can presently be compared.}

\subsection{The Arecibo Observatory 327-MHz Drift Survey} 
\begin{figure*}
    \centering
     \includegraphics[width=.9\linewidth]{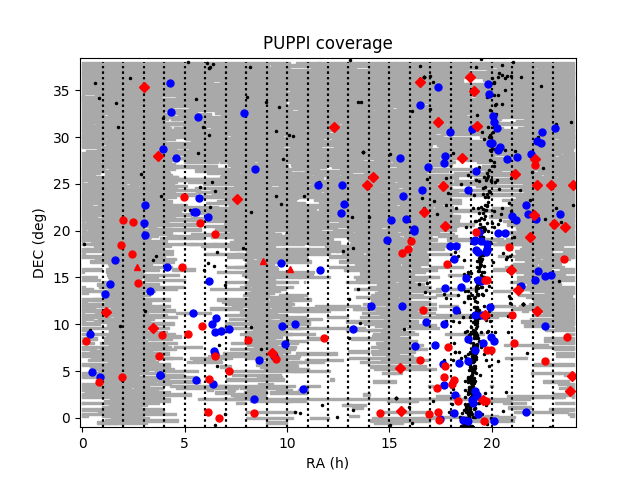}

    \caption{Both known and AO327-discovered  pulsars in the sky visible from Arecibo.  Grey strips indicate sky observed using the PUPPI backend. Black dots correspond to known pulsars from the ATNF catalog. Blue circles show known pulsars detected by the AO327 search pipeline. Red triangles indicate pulsars discovered using the WAPP backend, while red circles and red diamonds were found using the Mock and PUPPI backends.}
    \label{fig:sky}
\end{figure*}

The Arecibo Observatory (AO) 327-MHz Drift Survey \citep[AO327,][]{Deneva_2013} began in 2010 and ended in December 2020, making use of time when the 305-m telescope was not fully functional, such as during repairs or under tropical storm watch, as well as under-subscribed local sidereal time  ranges (LSTs). A “drift” scan is performed by positioning the telescope at a fixed azimuth and zenith angle, and allowing the sky to pass overhead. Due to the tragic collapse of Arecibo in December 2020, the AO327 survey ended before completing its intended sky coverage, with 65\% of the Arecibo sky observed, as shown in Fig.~\ref{fig:sky}.  The complete dataset has been processed through our search pipeline, with survey candidates being migrated to pulsars.nanograv.org for inspection by high-school and undergraduate students \citep{gateways}. To date, our survey has yielded 105 pulsar discoveries in total\footnote{\url{http://ao327.nanograv.org/newpulsars/}}, including the ones we report in this work and in previous publications (see Table~\ref{tab:discoveries}). Out of those, 16 are recycled pulsars \citep{Deneva_2013,Martinez_2017,Martinez_2019,Lewis_2023} and 19 are so-called RRATs, only detectable through single pulses \citep{Deneva_2013,Deneva_2016}. Of the 16 recycled pulsars, 10 are MSPs with three with such precise timing, they are now apart of the NANOGrav PTA\citep{Agazie_2023}.

\begin{table*}[]
    \centering
    {\tiny 
    \begin{tabular}{llllclll}
    \toprule
    \toprule
    Discovery name & Final name   & $P$    & DM             & Binary &	 Discovery date  &   Pipeline(s) 	& Reference \\ 
(J2000)       & (J2000)    & (s)    & (pc cm$^{-3}$) &         &	 (yyyy mm dd)       	&           \\ 
\toprule	 	
	 
J0011+08      & J0011+0805 & 2.55287 & 24.8           &        &	 2014 05 21    &   FFT, SP 	&  (6), (13)  \\ 
J0050+03      & J0050+0348 & 1.36656 & 26.4           &        &	 2014 05 22    &   FFT, SP	&  (6), (13)  \\
              & J0154+1833 & 0.00236 & 19.8        &         &	 2013 09 11    &   FFT          	&  (9)\\
J0156+04      &            &         & 27.5	          &         &	 2014 05 20    &   SP	&  (6)\\
J0158+21      & J0158+2106 & 0.50528 & 19.9           &        &	 2012 08 30    &   FFT             	&  (2), (13)  \\ 
               & J0225+1727 & 0.39031 & 20.0           &        &	 2013 09 15    &   FFT, SP  	& (13)    \\ 
J0229+20      & J0229+2058 & 0.80688 & 26.7           &        &	 2012 09 03    &   FFT, SP 	&  (2), (13)  \\ 
J0241+16      & J0241+1604 & 1.54530 & 19.6           &        &	 2005 04 03    &   FFT, SP	&  (2), (13)  \\ 
J0244+14      & J0245+1433 & 2.12748 & 29.5           &        &	 2012 08 10    &   FFT, SP  	&  (2), (13)  \\ 
% PF: this is a table of published pulsars, so the following entry should not be here, unless we published it somewhere. Unlikely, as it is not in the ATNF catalogue.
%              & J0354+0847 & 0.99888 & 66.0           &        &	 2015 04 30    &   FFT, SP	&   \\ 
J0453+16      & J0453+1559 & 0.04578  & 30.3        & yes      &	 2012 08 10    &   FFT            	& (2), (4)\\
J0457+23      & J0457+2333 & 0.50491 & 58.7           &        &	 2012 08 19    &   FFT, SP	&  (2), (13)  \\ 
J0509+08      & J0509+0856 & 0.00406 & 38.3         & yes      &	 2013 04 18    &   FFT	& (2), (9)\\
J0544+20      &            &         & 56.9	          &         &	 2014 02 07    &   SP	&  (6)\\
J0550+09      &            & 1.745   & 86.6	          &         &	 2014 10 15    &   SP	&  (6)\\
J0608+00      & J0608+0044 & 1.07619 & 48.4          &        &	 2012 10 06    &   FFT, SP	&  (2), (13)  \\ 
J0611+04      & J0611+0411 & 1.67443 & 69.6           &        &	 2014 02 03    &   FFT, SP	&  (6), (13)  \\ 
J0628+06      & J0627+0649 & 0.34652 & 86.5          &        &	 2012 11 26    &   FFT            	& (1)*, (13) \\ 
J0630+19      &            & 1.24855 & 48.3	          &        &	 2015 04 09    &   FFT, SP	&  (6)\\
              & J0639$-$0004 & 2.40949 & 70.1         &        &	 2013 01 22    &   FFT, SP	&  (13) \\ 
              & J0709+0458 & 0.03443 & 44.3        & yes    &	 2013 10 21    &   FFT            	&  (9)\\ 
              & J0732+2314 & 0.00409 & 44.7        & yes    &	 2016 10 07    &   FFT            	&  (9)\\
J0806+08      & J0806+0811 & 2.06310 & 46.7           &        &	 2013 06 18    &   FFT, SP	&  (2), (13)  \\ 
J0824+00      & J0824+0028 & 0.00986 & 34.5       & yes      &	 2012 10 27    &   FFT            	&  (2), (9)\\
J0848+16      & J0848+1640 & 0.45226 & 38.6           &        &	 2006 06 20    &   FFT, SP	&  (2), (13)  \\ 
              & J0916+0658 & 0.04477 & 19.2           &        &	 2017 06 22    &   FFT            	&  (13)  \\ 
J0928+06      & J0928+0614 & 2.06036 & 50.5           &        &	 2013 07 01    &   FFT, SP	&  (2), (13)  \\ 
J1010+15      &            &         & 42.2          &         &	 2012 10 27    &   SP	&  (2)\\
              & J1147+0829 & 1.62478 & 26.9           &        &	 2014 04 28    &   FFT, SP	&  (13) \\
              & J1215+3058 & 0.83596 & 15.6           &        &         2016 11 10    &   FFT, SP   & (13)   \\
              & J1411+2551 & 0.06245 & 12.4         & yes    &	 2014 09 09    &   FFT            	& (7)\\
J1433+00      &            &         & 23.6	          &         &	 2014 07 18    &   SP	&  (6)\\
              & J1531+0519 & 1.41982 & 31.3           &        &	 2017 04 27    &   FFT, SP	&  (13) \\ 
              & J1538+1736 & 0.69024 & 34.6           &        &	 2014 09 01    &   FFT            	&  (13)  \\ 
J1554+18      &            &         & 24.0	          &        &	 2014 09 13    &   SP	&  (6)\\
J1603+18      &            & 0.503   & 29.7	          &         &	 2014 09 16    &   SP	&  (6)\\
              & J1628+0613 & 1.67847 & 53.9	          &        &	 2013 10 20    &   FFT, SP	&  (13) \\ 
              & J1630+3550 & 0.00323 & 17.5	          & yes      &	 2018 01 17           &   FFT  &  (11)  \\ 
              & J1637+1131 & 1.67847 & 53.9	          &        &	 2013 11 27    &   FFT, SP		&  (13)  \\ 
J1656+00      & J1656+0018 & 1.49785 & 47.4	          &        &	 2014 08 05    &   FFT, SP	&  (6), (13)  \\ 
J1717+03      &            & 3.901   & 25.6	          &         &	 2014 01 22    &   SP	&  (6)\\
J1720+00      &            & 3.357   & 46.2	          &         &	 2014 05 20    &   SP	&  (6)\\
J1726$-$00    & J1726$-$0022 & 1.30862 & 59.9	      &        &	 2013 03 24    &   FFT            	&  (2), (13)  \\ 
J1738+04      & J1738+0418 & 1.39179 & 23.5           &        &	 2014 10 15    &   FFT, SP	&  (6), (13)  \\ 
% PF: the following pulsar was found independently at Puschino
              & J1742+2022 & 0.25258 & 19.8	          &        &	 2016 04 05    &   FFT, SP	&  (12)* \\ 
J1743+05      & J1743+0532 & 1.47364 & 55.4	          &        &	 2014 05 21    &   FFT, SP	&  (6), (13)  \\ 
J1749+16      & J1749+1629 & 2.31132 & 59.5	          &        &	 2014 10 23    &   FFT, SP	&  (6), (13)  \\ 
J1750+07      & J1750+0733 & 1.90881 & 55.7	          &        &	 2014 05 08    &   FFT, SP	&  (6), (13)  \\ 
J1802+03      & J1802+0344 & 0.66426 & 76.9	          &        &	 2013 01 13    &   FFT            	&  (2), (13)  \\ 
J1807+04      & J1807+0359 & 0.79885 & 52.7	      &        &	 2012 10 20    &   FFT, SP	&  (2), (13)  \\
J1821+01      & J1821+0155 & 0.03378 & 51.8           &        &	 2012 10 31    &   FFT            	&  (2), (3)* \\
              & J1832+2749 & 0.63170 & 47.4	          &        &	 2014 10 14    &   FFT            	&  (13) \\ 
% PF: The following "pulsar" has no associated data, nor in http://ao327.nanograv.org/newpulsars/, so I commented it out
%              & J1845+3543 &         &                &        &	               &              	&   \\ 
              & J1912+1947 & 2.37616 & 94.2	          &        &	 2014 07 16    &   FFT, SP	&  (13) \\ 
              & J1917+3115 & 1.84024 & 81.1	          &        &	 2016 11 11    &   FFT, SP	&  (13) \\ 
J1937$-$00    & J1937$-$0023 & 0.24015 & 67.9         &        &	 2013 07 11    &   FFT            	&  (2), (13)  \\ 
J1938+14      & J1938+1505 & 2.90251 & 74.4	          &        &	 2014 10 23    &   FFT, SP	&  (6), (13)  \\ 
J1941+01      & J1942+0147 & 1.40473 & 133.2          &        &	 2014 05 15    &   FFT, SP	&   (6), (13)  \\ 
J1945+07      & J1945+0720 & 1.07394 & 62.3	      &        &	 2012 10 11    &   FFT, SP	&  (2), (13)  \\ 
J1946+14      & J1946+1447 & 2.28244 & 50.5	          &        &	 2014 10 23    &   FFT, SP	&  (6), (13)  \\ 
J1956+07      & J1957+0724 & 5.01248 & 61.8           &        &	 2015 05 01    &   FFT, SP	&  (6), (13)  \\ 
              & J2050+1820 & 5.0483	 & 63.1	          &        &	 2013 09 10    &   FFT, SP	&  (13) \\ 
              & J2055+1545 & 0.00216 & 30.3	          & yes      &	 2016 02 16           &   FFT  &  (11)  \\ 
              & J2059+1100 & 0.95389 & 60.3	          &        &	 2014 10 10    &   FFT            	&  (13)  \\ 
J2105+07      & J2105+0757 & 3.74663 & 52.5	          &        &	 2014 09 15    &   FFT, SP	&  (6), (13)  \\ 
              & J2116+1345 & 0.00222 & 30.3	          & yes      &	2016 02 29          &   FFT  &  (11)  \\ 
% PF: the following pulsar was found independently at Puschino
              & J2151+1918 & 1.03372 & 30.9	          &        &	 2016 04 19    &   FFT, SP	&  (12)*, (13) \\ 
              & J2202+2134 & 1.35727 & 17.8	          &        &	 2017 05 01    &   FFT, SP	&  (8)*, (13) \\ 
J2204+27      & J2204+2700 & 0.08470 & 35.1         & yes      &	 2011 08 24    &   FFT            	& (2), (9)\\
              & J2212+2450 & 0.00391 & 30.3	          &      &	   2017 11 10       &   FFT  &  (11)  \\ 
J2234+06      & J2234+0611 & 0.00358 & 10.8         & yes      &	 2013 01 04    &   FFT          	& (2), (5), (10) \\
              & J2252+2455 & 1.79762 & 34.8	          &        &	 2017 09 18    &   FFT, SP	&   (13) \\ 
J2329+16      & J2329+1657 & 0.63207 & 30.4	          &        &	 2013 04 04    &   FFT, SP	& (2), (13)  \\ 
J2340+08      & J2340+0831 & 0.30330 & 23.8	      &        &	 2012 10 14    &   FFT, SP	& (2), (13)  \\ 
              & J2347+0300 & 1.38347 & 16.1           &        &	 2016 03 01    &   FFT, SP	&  (13) \\ 
% PF: the following pulsar was found independently at Puschino
              & J2354+0434 & 0.95635 & 12.6	          &        &	 2016 02 23    &   FFT, SP	& (12)*, (13)  \\  
 %    \bottomrule
    \end{tabular}
    \caption{{\footnotesize Published pulsar discoveries (including this work) from the AO327 survey. Column 1--2 indicate initial and, where applicable, final pulsar name from the timing solution localizations. Columns 3--5 list basic properties (spin period, DM, binary status). Column 6--7 indicates discovery dates and which pipelines yielded detections (FFT and or single-pulse). Lastly, column 8 provides the latest references on these pulsars. Those indicated with asterisks were independently co-discovered by other teams. References by date of publication are: (1) \citet{burgay2013}, (2) \citet{Deneva_2013}, (3) \citet{2013ApJ...768...85R}, (4) \citet{Martinez_2015}, (5) \citet{Antoniadis_2016}, (6) \citet{Deneva_2016}, (7) \citet{Martinez_2017}, (8) \citet{tyulbashev18}, (9) \citet{Martinez_2019}, (10) \citet{Stovall_2019}, (11) \citet{Lewis_2023},
    (12) \cite{tyulbashev24} and (13) this work.}} 
    
    \label{tab:discoveries}}
\end{table*}

The survey is also sensitive to emission from known pulsars and we actively maintain a catalog\footnote{\url{http://ao327.nanograv.org/}} of those detections \citep{catalog_paper}. These observations are valuable for studying pulse profile evolution, emission behaviors, scintillation, and other science. In the course of the survey, we have also developed two novel classifiers for single-pulse candidates, \textsc{Clusterrank}\footnote{\url{https://github.com/juliadeneva/clusterrank}} \citep{Deneva_2016} and \textsc{SPEGID}\footnote{\url{https://github.com/dipangwvu/SPEGID}} \citep{10.1093/mnras/sty1992}. A more recent re-searching of the AO327 dataset using Heimdall and FETCH \citep{fetch} is being undertaken as apart of The Petabyte project (TPP) \citep{lewis2023petabyteproject}.

In addition to facilitating studies of emission behaviors, dedicated follow-up observations for a year or more allow the determination of timing solutions which constrain astrometric and spin-down parameters. In this paper, we present new pulsar discoveries, timing solutions for both new and previously published discoveries, and a brief census of emission properties for 49 pulsars discovered in the AO327 survey. Of these, 31 pulsars have already been published, while 18 are new discoveries. One of the 49 pulsars is a partially recycled pulsar. These pulsars were observed between the periods of 2013 -- 2019 primarily using  AO's 327-MHz and L-band receivers. In \S\ref{observations} we describe our observations, \S\ref{search_pipeline} describes our survey's search pipeline,  \S\ref{timing} details our procedure for timing, \S\ref{results} describes our results, and \S\ref{discussion} gives a brief discussion of our results and conclusions.

\section{Observations} \label{observations}
\subsection{Survey Observations}
All sources presented in this paper were discovered with drift-scan observations at a center frequency of 327 MHz. The full width at half maximum power beam size of the Arecibo telescope at this frequency is 15$^\prime$, giving an approximate source transit time of about 1 minute. Raw total intensity data was taken and recorded using the ``incoherent search mode'' of the Puerto Rico Ultimate Pulsar Processing Instrument (PUPPI) backend at a sampling time of 81.92 $\mu$s.
The observations were configured to use 2816 frequency channels across the 68.75-MHz effective bandwidth of the 327-MHz receiver. Each full drift-scan observation was stored in a 4-bit format following the PSRfits convention\footnote{\url{https://www.atnf.csiro.au/research/pulsar/psrfits_definition/Psrfits.html}}. For more details regarding the survey design, please refer to \citet{Deneva_2013}.

\subsection{Search Pipeline} \label{search_pipeline}
 Discoveries were made using our \textsc{presto}\footnote{\url{https://github.com/scottransom/presto}}-based search pipeline \citep{rem02,ar18,ran11}. Search observations were first divided into 60-sec intervals with a stride of 30 sec (i.e., each interval overlaps by 30 sec of the adjacent interval) so as to ensure full coverage of sources transiting across the telescope beam. The  search pipeline was then deployed on each 60-sec interval. Using the \textsc{presto} routine, \texttt{rfifind}, radio frequency interference (RFI) mitigation was performed by inspecting time chunks of $\sim$2 seconds and flagging outlier time and frequency chunks corrupted by RFI. For more details of \texttt{rfifind} functionality, we refer the reader to \cite{accel_sift}.

After RFI mitigation, the next steps are to remove the effects of
dispersion by ionized gas along the line of sight (``dedispersion") by
summing the time series over all frequency channels, shifting each by the time needed to correct for dispersion delay,  so as to
maximize signal strength in the time domain. Interstellar dispersion
is well-studied and imparts a frequency-dependent delay of
 \begin{equation}
 \label{dm_delay}
    \Delta t = t_{2} - t_{1} = 4.15 \times {\rm DM} \times (\nu_{1}^{-2}-\nu_{2}^{-2})
 \end{equation}
where DM is the dispersion measure in units of pc cm$^{-3}$, $\nu{1}$ and ${\nu_{2}}$ are the center frequencies of two channels given in GHz, and the delay between the arrival times in two channels $\Delta t$ is measured in ms. In practice one defines the DM as the free electron column density, though there are much smaller contributions from ions and other effects along the line of sight \citep{kulkarni2020dispersionmeasureconfusionconstants}. Our maximum DM is set to correspond to a dispersion delay equal to the transit time of a source through the 327-MHz beam, while the DM step size is optimized based on the observing frequency, bandwidth and sampling time. We use the \textsc{presto} routine \texttt{DDPlan.py} to compute the dedispersion plan as well as appropriate downsampling. Dedispersion was then performed utilizing \texttt{prepsubband} with downsampling factors of 1, 2, 4, and 8. Our maximum DM of 1438.2 pc~${\rm cm}^{-3}$ is much higher than the expected Galactic contributions along all of the lines of sight sampled in our survey, providing good sensitivity to extragalactic FRBs.

 There are two standard approaches for identifying astrophysical sources in dedispersed time series. The first uses Fourier analyses to detect periodic signals. These include the Fast Fourier Transform aided by harmonic stacking, and acceleration searches to increase sensitivity to binary pulsars \citep{2001PhDT.......123R}. The apparent spin periods of pulsars in binary systems will vary due to Doppler shifts from binary accelerations. The degree of acceleration can be captured using parameter $z$, the Fourier frequency derivative that is induced by the observed variation in the rotational period. Standard convention is to define $z$ in units of frequency bins drifted per observation length. We perform both low acceleration searches ($z_{max}=0$), and high acceleration searches ($z_{max}=50$). Harmonic summing was performed up to harmonics of order 16 (8) in our low (high) acceleration searches. We then used \texttt{ACCELSift.py}\footnote{A description of the code can be found in Section 3.3.4 of \citet{accel_sift}.} to filter out spurious candidates caused by RFI or degeneracies in the Fourier search space. Candidates were then posted to the Pulsar Science Collaboratory pulsar searching website\footnote{\url{http://pulsars.nanograv.org}} for visual candidate inspection \citep{gateways}.

The second method is geared towards astrophysical sources that either are non-periodic or do not exhibit easily identifiable periodicity in their emission. Single-pulse searches \citep{2003ApJ...596.1142C} exploit the dependence of signal-to-noise that an astrophysical pulse exhibits in the time vs DM parameter space. For each trial DM, downsampling is applied through convolution of a boxcar function so as to maximize signal to noise of any candidate single pulses. We perform a single-pulse search  with PRESTO's \texttt{single\_pulse\_search.py} using boxcar widths ranging from 1 to 30 samples, corresponding to roughly  0.1 s, and choosing candidate pulses with a signal-to-noise ratio of 5.5$\sigma$ or higher to be flagged for further inspection. We then applied two separate classifiers, \textsc{Clusterrank} and \textsc{SPEGID}, to filter out the best single-pulse candidates.

\subsection{Timing Observations}\label{timing_obs} 
Following discovery, a timing campaign was undertaken using AO's 327-MHz and L-band receivers. Primary observations utilized the 327-MHz receiver in conjunction with the PUPPI backend. Observations varied in length, with a minimum integration time of 100 seconds in order to ensure a stable integrated profile. For each source, roughly half of the timing observations were taken in PUPPI's coherent search mode and the remaining in coherent fold mode. The coherent modes perform coherent dedispersion, which more accurately removes the effects of interstellar dispersion compared to incoherent dedispersion \citep{coherent_disp}, and search mode allowed for at least half of the observations for each source to have detectable single-pulses. Discovery dispersion measures (DMs -- see Sec \ref{search_pipeline} for details) and the drift-scan backend configurations were used for these initial observations. We used 128 channels in coherent search mode. Once a satisfactory preliminary timing solution was found, we then used PUPPI's coherent fold mode configured for full Stokes with 56 to 256 frequency channels. Sub-integration lengths of $\sim$10.5 seconds were used for all sources except for J0916+0658. Sub-integration lengths must be shorter than the time it takes for profile drift caused by an incorrect folding period to become significant, which roughly scales with a pulsar's period. As J0916+0658 has a period of $\sim$45 ms, we used a sub-integration length of $\sim$2.5 seconds. Calibration data taken before each fold-mode observation consisted of an injection of a 25-Hz linearly polarized signal using a noise diode. Calibrator observations were 60 seconds, where 50\% of this time had the noise diode engaged. Observations were taken over an approximately six-year period. A program of several closely spaced observations followed or preceded by a number of observations of varying cadence is generally sufficient for constraining the spin and astrometric parameters to a sufficient degree of precision. We observed each pulsar  at roughly daily cadence for several days to establish initial phase connection while remaining observations were taken at roughly weekly and monthly cadences depending on availability.  In cases where a source was weakly detectable at L-band, at most a handful of L-band observations were taken at a roughly weekly or monthly cadence while bright L-band sources also included a handful at daily cadence.

Observations using the L-band receiver were conducted for a subset of the 49 pulsars in conjunction with the PUPPI backend. All L-band observations were taken using PUPPI's coherent search mode using a center frequency of 1380 MHz,  a sampling time of 40.96 $\mu$s, and 2048 frequency channels across the 800-MHz effective bandwidth. After coherent dedispersion within each of the 2048 channels, the frequency resolution was decimated to 128 channels. We first conducted a single observation of each source to estimate its L-band signal-to-noise. Bright sources were prioritized for further observations, while weaker sources were observed no more than a few epochs. We carried out no follow-up observations for non-detected sources.

\section{Analysis} \label{timing}
After we confirm the discovery of a pulsar, we carry out a dedicated follow-up campaign (see Sec \ref{timing_obs}) to obtain its phase-connected timing solution. The data is also used to estimate some of their useful physical parameters such as characteristic age and inferred surface magnetic field strength \citep{2004hpa..book.....L}. The details of these analyses are given below.

\subsection{Timing Analysis}

\begin{figure*} 
    \figurenum{2.1}
    \plotone{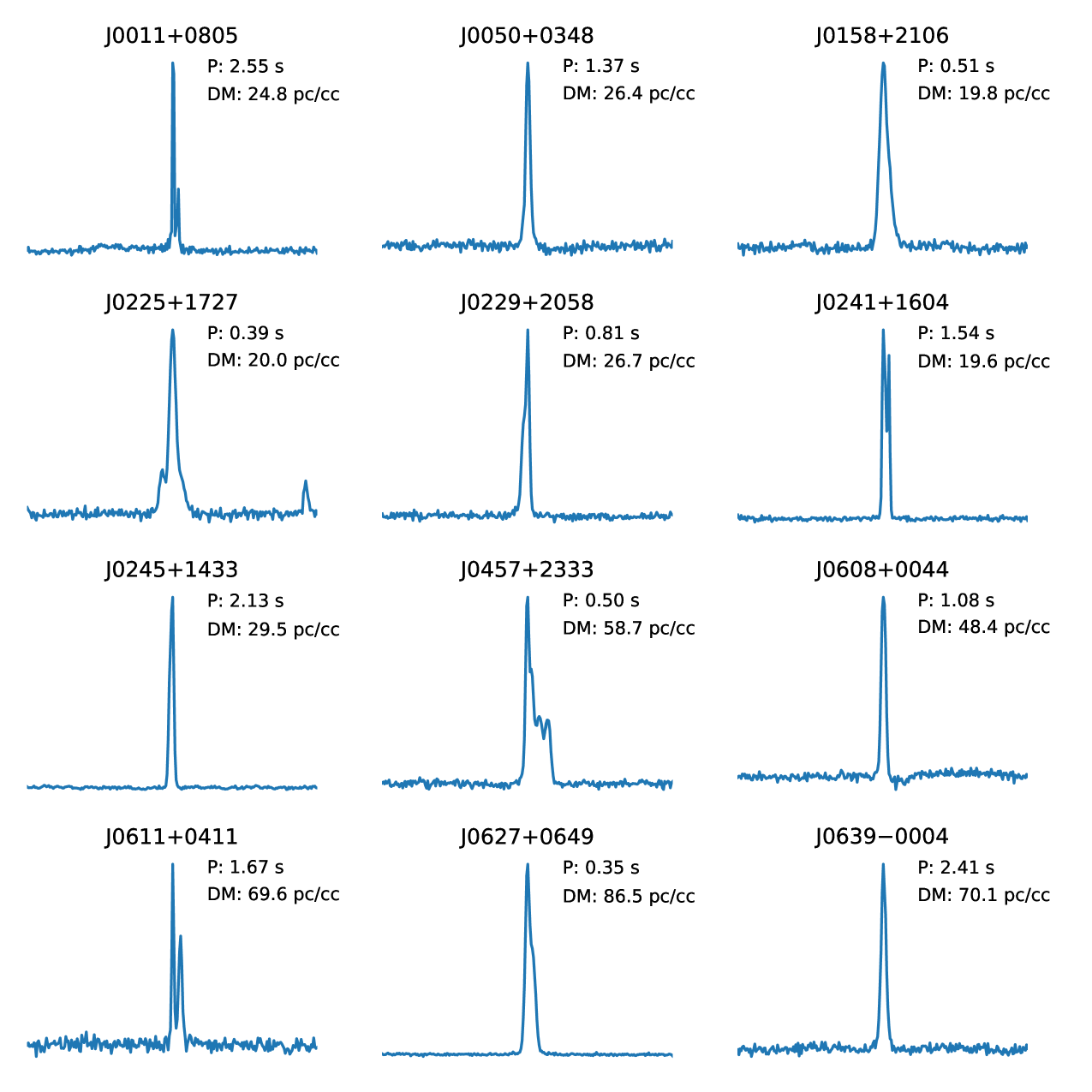}
    \caption{Epoch-averaged intensity profiles at 327 MHz (blue) and 1380 MHz (red). We plot intensity on the y-axis and rotational phase on the x-axis. One full rotation is shown. Each profile has  a resolution of 256 phase bins. We label each profile with the  pulsar's period and DM. The complete figure set (6 images) is available in the online journal.}
    \label{fig_avp1}
\end{figure*}

 A time of arrival (TOA) is the absolute time at which a pulse is detected as measured at the observatory. We  formed period-averaged profiles for each  epoch of data using 256 bins, and  measured TOAs by convolving this profile with a high signal-to-noise template profile. To create templates, we first excluded observations where RFI was egregious or the S/N was too low. We then selected an arbitrary epoch to serve as the reference profile and made use of the `Fourier-gradient method' \citep{1992RSPTA.341..117T} to phase align each epoch's profile with the reference profile. The phase offset between each individual and reference profile was computed by taking the Fourier transform of both and performing a cross-correlation. All profiles were then averaged to form an epoch-averaged profile of 256 bins. These are shown in Fig~2.1 --~2.6. The epoch-averaged profiles were then used as our templates for timing at 327 MHz. Given the spectra of these pulsars, noise-free templates were only merited at L-band. For L-band observations, we utilized noise-free Gaussian templates to reduce uncertainty in our TOA measurements. 

To measure TOAs, for most observations we divided the complete observation into two subintegrations and two subbands, resulting in four TOAs per observation. For observations in coherent search mode, we used the PRESTO routine \texttt{get\_TOAs.py} while those in fold mode we used PSRCHIVE's \texttt{pat}. For each TOA, we measure a corresponding uncertainty in the arrival time. 

TEMPO2 \citep{tempo2} was utilized to construct the timing solutions, and unless otherwise noted, we use its default configuration and conventions. 
This code first converts TOAs from observatory topocentric time to Barycentric Coordinate Time (TCB)  with the use of JPL's DE436 ephemeris \citep{de436_disc,de436}. We  fit the TOAs to a timing model accounting for spin period, period derivative, position, and dispersion measure through least-squares minimization of the residuals, or the differences between the measured and timing model-predicted TOAs. Depending on the complexity of the model, a parameter is added, and the process is reiterated for the remaining number of parameters, until all TOAs are phase-connected with low co-variances in the model parameters or relatively flat residuals on long time-scales. Because the uncertainties in the TOAs influence the uncertainties of parameters in the timing model, it is important to ensure that the  errors on timing model parameters are not underestimated \citep{10.1093/mnras/stw347}. It is convention to assume that the true uncertainty for a given TOA ($\sigma_{new}$) is simply proportional to its measured value ($\sigma_{old}$), $\sigma_{new}=F \sigma_{old}$, where the system-dependent constant of proportionality, $F$, is usually referred to as the EFAC. We then assume that the model is a perfect fit to the data and therefore scale the uncertainty such that the reduced chi-squared value of the fit, $\chi_{\mu}^{2}=1$.

For all pulsars, we were able to construct phase-connected solutions using only basic astrometric and spin parameters and DM. Timing models, or .par files, are available for direct download via this paper's github site\footnote{ \url{https://github.com/tolszans/AO327/}}.  PUPPI's coherent search and coherent fold modes have different clock offsets, resulting in an intrinsic offset between the two groups of TOAs. We accounted for this  by fitting a phase jump in the timing model when appropriate. Table~\ref{table1} lists the fitted parameters as well as their associated errors. All solutions result in weighted root-mean-square residuals less than 0.3\% of their spin periods. Furthermore, they also result in  errors of less than 1\% in their frequency derivatives. All solutions are adjusted to a reference epoch corresponding to the center of our data-span.

\subsection{Polarization Calibration}
Polarization calibration was undertaken for a single source, PSR J0225+1727. We choose this source since interpulses arise from rare viewing geometries and polarization can be insightful in constraining the geometry. Fold-mode observations were calibrated using PSRCHIVE's \texttt{pac} and the single-axis (ideal feed) model \citep{loki_paper}. Because of Faraday rotation in the ISM, the polarized position angle (PPA) varies as a function of frequency \citep{2004hpa..book.....L}, which can lead to depolarization during frequency integration. Hence, we fit for a rotation measure using PSRCHIVE's \texttt{rmfit}.  We conducted the search over RMs ranging from --200 to +200 rad m$^{-2}$, using \texttt{rmfit}'s iterative position angle refinement to find the RM at which the flux of the linear polarization peaks. We note the errors reported by \texttt{rmfit} likely underestimate the true error. Lastly, in some observations, we note the signs of the PPA track and Stokes-V are flipped. This likely was a result of a change in feed convention, which was often unrecorded at Arecibo and has been noted by other authors \citep{Fiore_2025}. To correct for this, we chose one of the two signs as correct, and flipped the signs of Stokes-U and Stoke-V for the incident observations.

\subsection{Flux Measurements}
Flux densities were estimated from search mode observations through use of the radiometer equation\footnote{Using Equation 3 from \cite{catalog_paper}}. To account for the variable sky temperature, we followed the procedure outlined in \cite{catalog_paper} to estimate  temperatures  at 327 MHz. For the 327 MHz receiver, the system temperature was well-controlled and measured daily, with an average $T_{rec} = 113$~K. Prior to hurricane Maria's landfall on MJD 58012, the gain for the 327-MHz receiver was approximately 10 K/Jy, while after, the gain dropped to 8 K/Jy. 

L-band observations were taken before hurricane Maria. The mean gain of the L-band receiver was around 10.3 K/Jy while the system temperature was approximately 33 K. Estimation of the L-band sky temperature were made using the map available from the LAB HI survey \citep{lband_map,lband2}.

A flux density was estimated for each epoch and epoch-averaged flux densities are reported in Table~\ref{table1} along with  standard deviations.

\subsection{Distance Measurements}
Distances were estimated using DM measurements with the python package PyGEDM \citep{pygedm} for the NE2001 \citep{2002astro.ph..7156C} and YMW2016 \citep{2017ApJ...835...29Y} Galactic electron distribution models while the python package MWPROP was used for the newest model, NE2025 \citep{ocker2026ne2025updatedelectrondensity}. The distances are reported in Table~\ref{table1}. In cases where a source's DM exceeds a given model's maximum DM,  we do not report a distance estimate. Healpy \citep{hp1,hp2}, the python implementation of HEALPix\footnote{healpix.sourceforge.net}, was used for plotting purposes.

\section{Results} \label{results}
We present the first phase-connected timing solutions for 49 pulsars discovered in the AO327 survey. Of these, 18  are new discoveries. All but one of our sources appear to be non-recycled, or slow, pulsars.

\subsection{Timing Results}
\setcounter{figure}{3}
\renewcommand{\thefigure}{\arabic{figure}.1}

\begin{figure*}
\begin{tabularx}{\textwidth}{Y}
\figurenum{3.1}
\includegraphics[page=1,angle=0.,width=1.\linewidth]{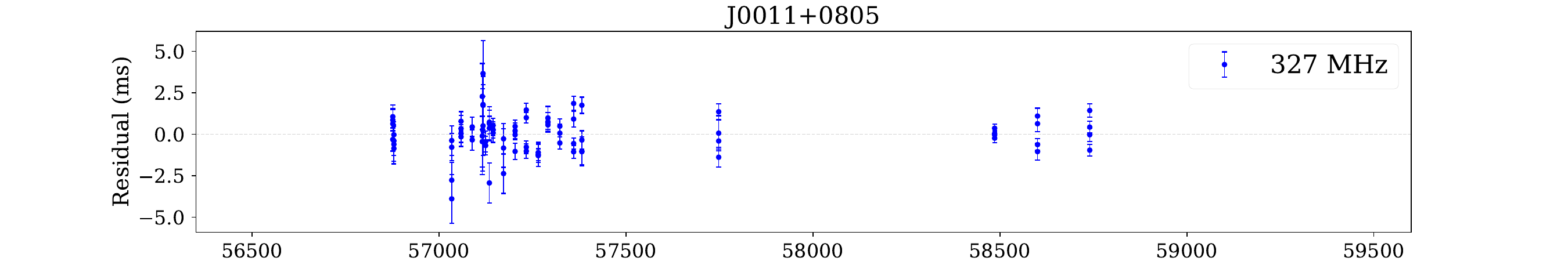} \\
\includegraphics[page=1,angle=0.,width=1.\linewidth]{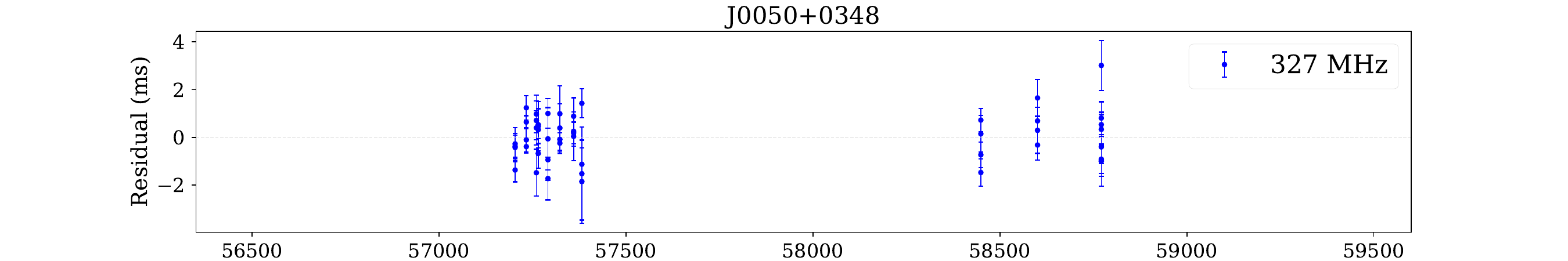} \\
\includegraphics[page=1,angle=0.,width=1.\linewidth]{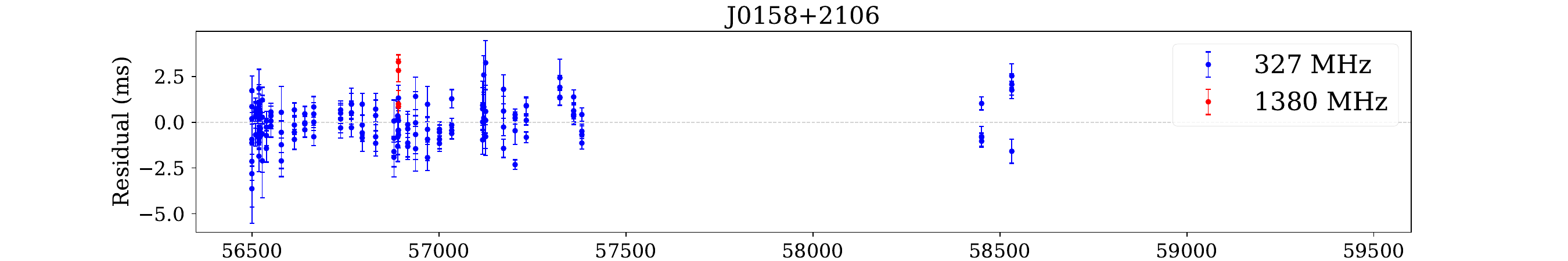} \\
\includegraphics[page=1,angle=0.,width=1.\linewidth]{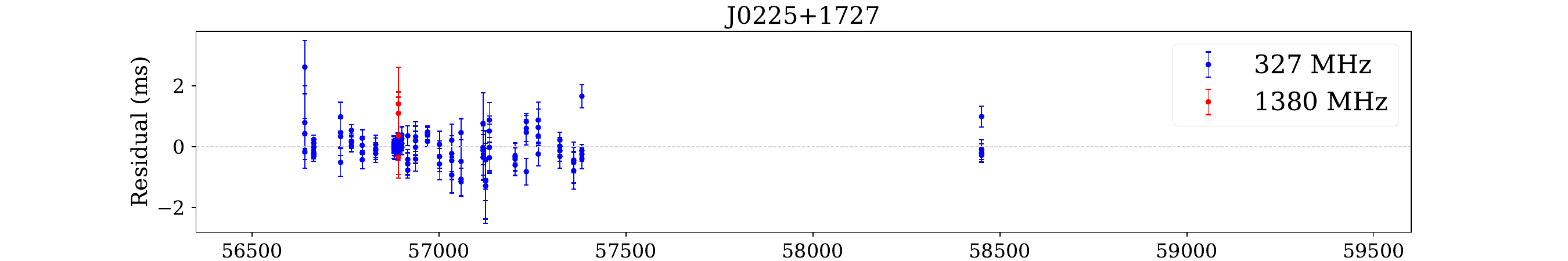} \\
\includegraphics[page=1,angle=0.,width=1.\linewidth]{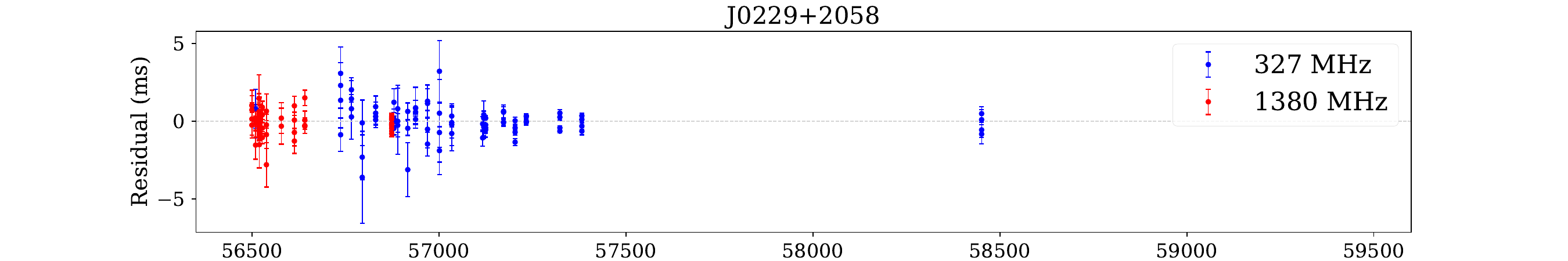} \\
\includegraphics[page=1,angle=0.,width=1.\linewidth]{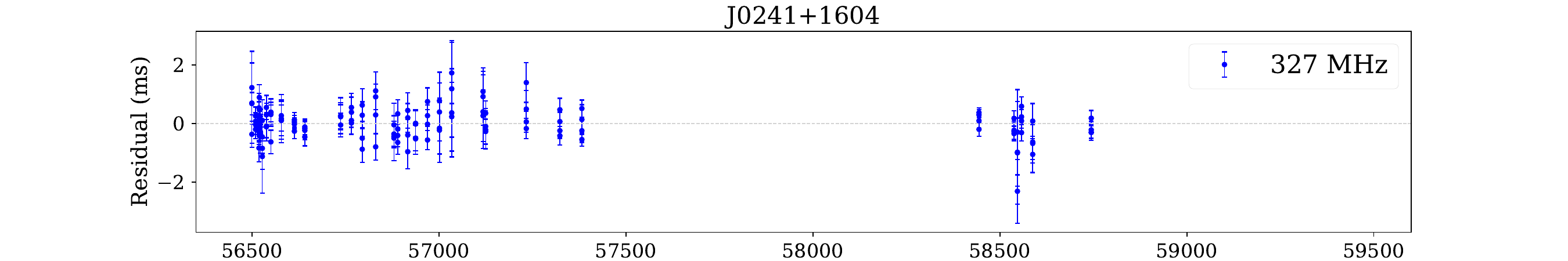} \\
\includegraphics[page=1,angle=0.,width=1.\linewidth]{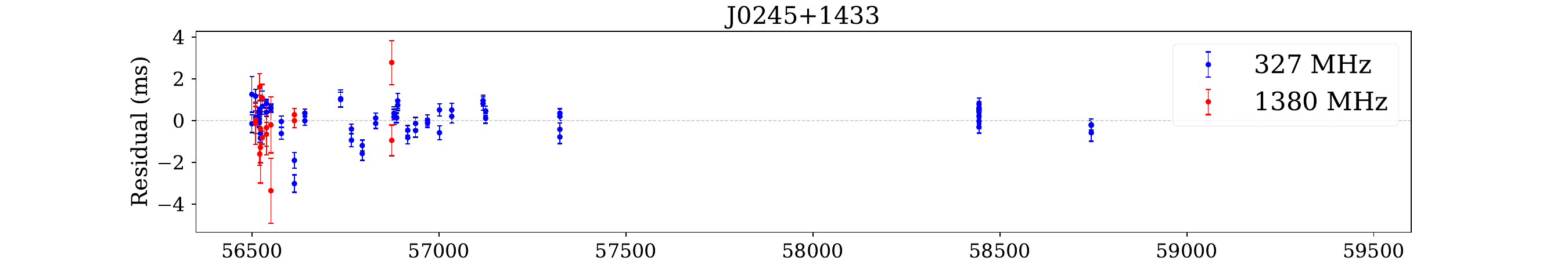} \\
\end{tabularx}
\caption{Residuals for individual timing solutions. We plot the postfit residual on the y-axis and the MJD on the x-axis. Blue coloring corresponds to data taken at 327 MHz while red coloring is used for 1380 MHz observations. Information regarding TOAs including  RMS and timespan can be found in Table~\ref{table1}.  The complete figure set (7 images) is available in the online journal.}
\end{figure*}

\renewcommand{\thefigure}{\arabic{figure}}

Timing residuals for each source are shown in Fig.~4.1 --~4.7. Of the 105 sources discovered in the AO327 survey, we present the timing solutions for 49 of these sources. Each timing solution includes constraints on basic astrometric and spin parameters. In most cases, we have updated the source names based on better localization via timing as described in Table~\ref{tab:discoveries}. 

Pulsar periods and period derivatives are compared to other populations in Fig.~\ref{ppdot}. The spin periods of our discoveries range from 40 ms to 5.05 s, and their DMs from 17.8 pc cm$^{-3}$ to 133.2 pc $\text{cm}^{-3}$. As evident in Fig.~\ref{ppdot}, we find that the majority of our discoveries skew towards older characteristic ages, as expected given the predominantly high Galactic latitudes it covered. This is consistent with other high Galactic latitude surveys such as the LOFAR Tied-Array All-Sky Survey \citep[LOTASS,][]{lotass}, the Green Bank North Celestial Cap survey \citep[GBNCC,][]{McEwen_2024}, and the GMRT High Resolution Southern Sky survey \citep[GHRSS,][]{Bhattacharyya_2019}.

\setcounter{figure}{3}
\begin{figure*}
    \centering
    \includegraphics[page=1,width=0.8\textwidth]{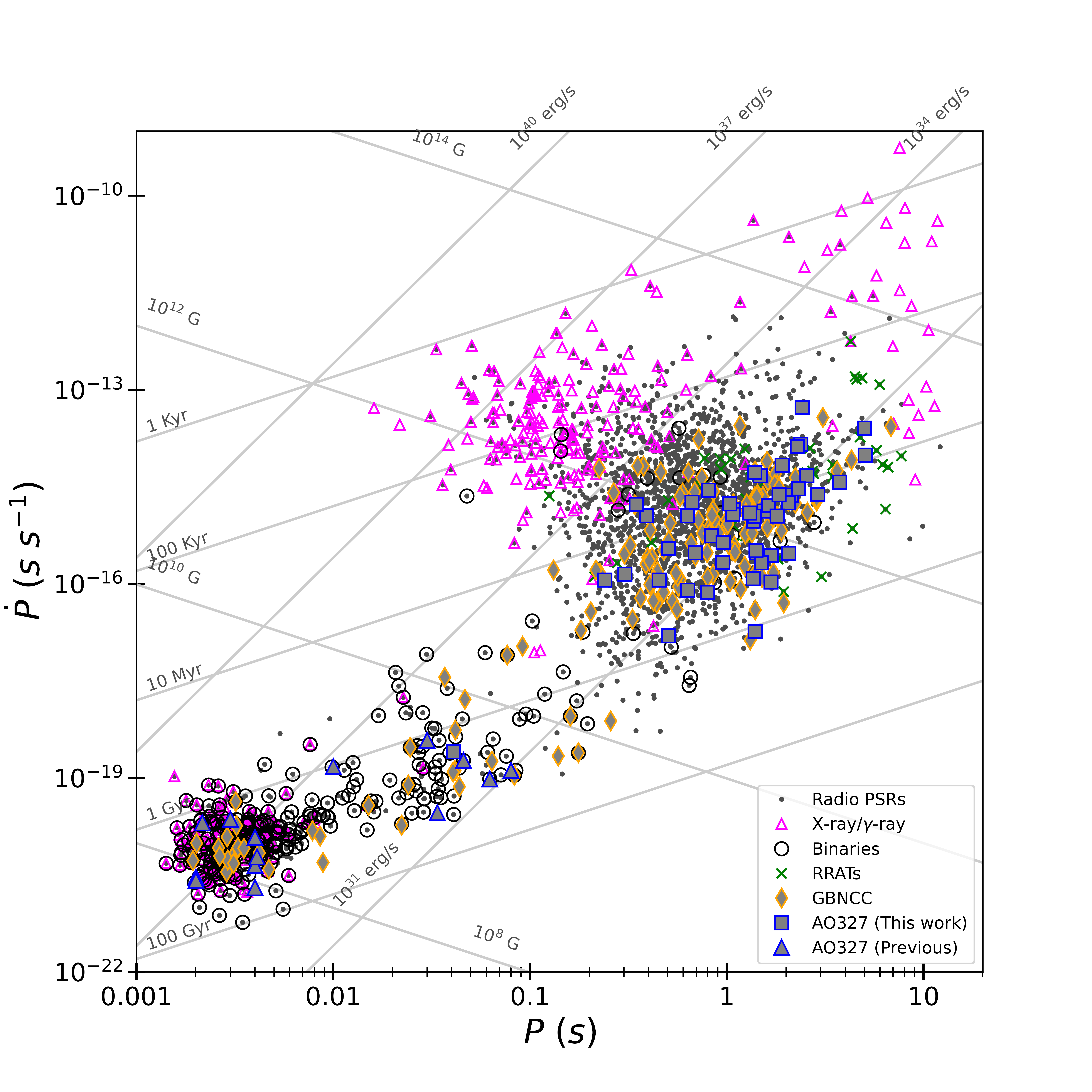}
    \caption{{\it P}-$\mathrm{\dot{\it P}}$ diagram showing the periods and period derivatives of known neutron stars. Black dots denote radio pulsars, purple diamonds X-ray/$\gamma$-ray pulsars, black circles systems with companion objects, green X's RRATs, gold diamonds pulsars discovered by GBNCC, blue squares AO327 pulsars being presented in this work, while blue triangles denote AO327 pulsars with previously published timing solutions. Also plotted are lines of constant $\dot{E}$, inferred magnetic field $B$, and characteristic age $\tau$. The lower left population comprises the recycled population, while the center population comprises non-recycled pulsars. We note that the longest period pulsars lie to the upper right of the population corresponding to higher magnetic fields while the middle and shortest period pulsars show a broader variation in characteristic ages and $\dot{E}$ values. Likewise, while AO327 has half as many discoveries as GBNCC, the discoveries from the two surveys mostly overlap in spin parameters. 
    %Clearer distinctions between surveys may become more apparent with more discoveries. 
    AO327 is expected to at least double its discoveries upon completion of candidate inspection.}
    \label{ppdot}
\end{figure*}

\newpage
\begin{sidewaystable}
    \tiny
    \setlength{\tabcolsep}{1pt}
    \vspace*{18cm}
    \hspace*{-4cm}
    \centering
    \begin{tabular}{lclllllcccccccccccccc}
    \toprule
    \toprule
    PSR & \multicolumn{1}{c}{R.A. (J2000)} & \multicolumn{1}{c}{Dec. (J2000)} & \multicolumn{1}{c}{$\nu$} & \multicolumn{1}{c}{$\dot{\nu}$} & \multicolumn{1}{c}{DM} & \multicolumn{1}{c}{Epoch} & \multicolumn{1}{c}{$N_{TOAs}$} & \multicolumn{1}{c}{RMS} & \multicolumn{1}{c}{$\chi^{2}$} & \multicolumn{1}{c}{Data Span} & \multicolumn{1}{c}{MJD Range} & \multicolumn{1}{c}{log($\tau$)} & \multicolumn{1}{c}{log($B_{surf}$)} & \multicolumn{1}{c}{log($\dot{E}$)} & \multicolumn{1}{c}{$D_{\rm NE2001}$} &  \multicolumn{1}{c}{$D_{\rm NE2025}$} & \multicolumn{1}{c}{$D_{\rm YMW16}$} & \multicolumn{1}{c}{$S_{p}$ $\pm$  $\sigma_{S_{p}}$} & \multicolumn{1}{c}{$S_{l}$ $\pm$ $\sigma_{S_{l}}$} & EFs  \\\\
(J2000) & \multicolumn{1}{c}{(h:m:s)} & \multicolumn{1}{c}{(d:m:s)} & \multicolumn{1}{c}{(Hz)} & \multicolumn{1}{c}{($ \text{ }Hz\text{ }s^{-1}$)} & \multicolumn{1}{c}{(pc cm$^{-3}$)} & \multicolumn{1}{c}{(MJD)} &  \multicolumn{1}{c}{}  & \multicolumn{1}{c}{($\mu$s)} & \multicolumn{1}{c}{}    &  \multicolumn{1}{c}{(Yr)} & \multicolumn{1}{c}{(MJD)} & \multicolumn{1}{c}{(Yr)}  &  \multicolumn{1}{c}{(G)}  & \multicolumn{1}{c}{(erg $s^{-1}$)} & \multicolumn{1}{c}{(kpc)} & \multicolumn{1}{c}{(kpc)} & \multicolumn{1}{c}{(kpc)} & \multicolumn{1}{c}{(mJy)} & \multicolumn{1}{c}{(mJy)} &  \\\\
\toprule
J0011+0805 & 00:11:42.32(3) & +08:05:42.6(10) & 0.3917159530882(20) & $-$7.2027(12)E-16 & 24.758(18) & 57808 & 81 & 823 & 3.0 & 5.1 & 56876---58740 & 6.9 & 12.5 & 31.0 & 1.3 & 2.1 & 5.0 & 0.3 $\pm$  0.1 &  & AM, M?, N \\
J0050+0347 & 00:50:23.80(15) & +03:47:58(5) & 0.7317672555072(20) & $-$5.005(4)E-16 & 26.42(3) & 57987 & 48 & 797 & 2.0 & 4.3 & 57203---58771 & 7.4 & 12.1 & 31.2 & 1.4 & 2.6 & < 25 & 0.4 $\pm$  0.3 &  & SR \\
J0158+2106 & 01:58:45.984(11) & +21:06:47.5(5) & 1.979030408151(7) & $-$1.37960(15)E-15 & 19.777(12) & 57515 & 160 & 930 & 3.8 & 5.6 & 56499---58532 & 7.4 & 11.6 & 32.0 & 0.9 & 1.4 & 1.5 & 0.7 $\pm$  0.3 & 0.31  & SR \\
J0225+1727 & 02:25:22.288(11) & +17:27:44.1(5) & 2.562757254921(5) & $-$7.44343(14)E-15 & 19.975(6) & 57546 & 128 & 288 & 1.6 & 5.0 & 56641---58451 & 6.7 & 11.8 & 32.9 & 0.8 & 1.3 & 1.4 & 0.8 $\pm$  0.4 & 0.17  & IP, SR \\
J0229+2058 & 02:29:11.995(9) & +20:58:34.3(4) & 1.239438282432(5) & $-$4.31612(9)E-15 & 26.664(6) & 57475 & 135 & 486 & 2.6 & 5.3 & 56500---58451 & 6.7 & 12.2 & 32.3 & 1.1 & 1.9 & 2.0 & 0.3 $\pm$  0.5 & 0.03 $\pm$ 0.02 & D?, M?, N, SR \\
J0241+1604 & 02:41:25.79(8) & +16:04:14(4) & 0.6472955962072(8) & $-$5.60985(20)E-16 & 19.558(7) & 57621 & 140 & 377 & 1.0 & 6.2 & 56499---58744 & 7.3 & 12.2 & 31.2 & 0.8 & 1.3 & 1.3 & 0.6 $\pm$  0.2 &  & AM \\
J0245+1433 & 02:45:19.89(6) & +14:33:19(3) & 0.4699470190362(13) & $-$5.2125(3)E-16 & 29.506(9) & 57621 & 86 & 598 & 5.4 & 6.2 & 56499---58744 & 7.2 & 12.4 & 31.0 & 1.2 & 2.3 & 2.5 & 1.0 $\pm$  0.5 &  & AM, N \\
J0457+2333 & 04:57:03.00(2) & +23:33:54(3) & 1.980583234463(4) & $-$6.170(16)E-17 & 58.672(7) & 57816 & 48 & 237 & 2.5 & 5.1 & 56890---58744 & 8.7 & 11.0 & 30.7 & 1.9 & 2.1 & 1.5 & 2.0 $\pm$  0.4 &  & AM, N \\
J0608+0044 & 06:08:48.540(3) & +00:44:11.25(9) & 0.929218888961(6) & $-$1.2149(5)E-15 & 48.434(9) & 56889 & 82 & 318 & 3.4 & 2.0 & 56522---57258 & 7.1 & 12.1 & 31.6 & 1.9 & 2.0 & 1.3 & 1.5 $\pm$  0.6 & 0.13 $\pm$ 0.12 & D \\
J0611+0411 & 06:11:51.800(18) & +04:11:09.8(3) & 0.597211849968(19) & $-$3.81(13)E-17 & 69.63(3) & 57308 & 47 & 867 & 1.0 & 2.2 & 56916---57701 & 8.4 & 11.6 & 30.0 & 2.4 & 2.3 & 1.7 & 0.2 $\pm$  0.1 &  & AM, D, N \\
J0627+0649 & 06:27:53.6275(7) & +06:49:54.06(4) & 2.885811444479(3) & $-$1.414561(6)E-14 & 86.512(4) & 57325 & 116 & 144 & 2.1 & 4.5 & 56500---58152 & 6.5 & 11.9 & 33.2 & 2.6 & 2.3 & 1.8 & 1.9 $\pm$  1.8 & 0.06 $\pm$ 0.06 & AM \\
J0639$-$0004 & 06:39:44.592(14) & $-$00:04:53.4(3) & 0.415009456624(9) & $-$9.2020(10)E-15 & 70.09(4) & 56950 & 75 & 1223 & 0.8 & 2.2 & 56554---57347 & 5.8 & 13.1 & 32.2 & 2.5 & 2.3 & 1.8 & 0.4 $\pm$  0.2 &  & AM, N \\
J0806+0811 & 08:06:18.629(7) & +08:11:54.8(4) & 0.484706968755(6) & $-$6.968(12)E-17 & 46.700(16) & 57396 & 150 & 930 & 3.2 & 4.9 & 56500---58293 & 8.0 & 11.9 & 30.1 & 2.0 & 3.3 & 1.9 & 0.6 $\pm$  0.2 & 0.42  & AM, M, N \\
J0848+1640 & 08:48:44.016(9) & +16:40:24.6(5) & 2.2106472850544(14) & $-$5.6118(3)E-16 & 38.5556(20) & 57366 & 152 & 112 & 1.1 & 4.8 & 56500---58234 & 7.8 & 11.4 & 31.7 & 1.7 & 3.6 & 2.4 & 0.6 $\pm$  0.2 & 0.04 $\pm$ 0.03 & AM, D?, M, N \\
J0916+0658 & 09:16:15.2394(17) & +06:58:32.89(9) & 24.52432584050(9) & $-$1.52(6)E-16 & 19.1720(16) & 58210 & 116 & 99 & 4.6 & 1.4 & 57957---58463 & 9.4 & 9.5 & 32.2 & 0.8 & 1.3 & 0.9 & 1.8 $\pm$  0.8 &  & \\
J0928+0614 & 09:28:29.908(3) & +06:14:08.43(10) & 0.4853784462631(6) & $-$4.21575(12)E-16 & 50.480(5) & 57481 & 194 & 270 & 1.7 & 5.4 & 56500---58463 & 7.3 & 12.3 & 30.9 & < 25 & < 25 & < 25 & 0.8 $\pm$  0.3 &  & AM \\
J1147+0829 & 11:47:42.756(19) & +08:29:04.0(7) & 0.615361380485(4) & $-$6.158(3)E-16 & 26.895(16) & 57644 & 113 & 851 & 1.4 & 3.3 & 57049---58240 & 7.2 & 12.2 & 31.2 & 1.6 & 3.5 & < 25  & 0.2 $\pm$  0.1 & 0.11 $\pm$ 0.07 & N \\
J1215+3058 & 12:15:58.977(8) & +30:58:45.97(19) & 1.196223223417(15) & $-$7.87(3)E-16 & 15.557(9) & 57999 & 60 & 362 & 1.3 & 1.1 & 57789---58210 & 7.4 & 11.8 & 31.6 & 1.3 & 1.7 & 1.9 & 0.3 $\pm$  0.2 &  & AM, N, SR \\
J1531+0519 & 15:31:40.35(3) & +05:19:49.3(6) & 0.704213869107(19) & $-$1.50(4)E-16 & 31.27(7) & 58444 & 63 & 2089 & 1.3 & 1.6 & 58155---58734 & 7.9 & 11.8 & 30.6 & 3.1 & 3.7 & < 25 & 0.5 $\pm$  0.2 &  & AM, M, N \\
J1538+1736 & 15:38:10.473(4) & +17:36:10.24(8) & 1.448665690319(6) & $-$6.331(3)E-16 & 34.588(13) & 57706 & 158 & 904 & 3.8 & 4.2 & 56936---58477 & 7.6 & 11.7 & 31.6 & < 25 & < 25 & < 25 & 0.3 $\pm$  0.2 & 0.10 $\pm$ 0.05 & \\
J1628+0613 & 16:28:53.936(5) & +06:13:54.60(14) & 0.723782796197(3) & $-$5.8215(10)E-16 & 51.701(7) & 57485 & 125 & 697 & 1.1 & 4.6 & 56647---58325 & 7.3 & 12.1 & 31.2 & < 25 & < 25 & < 25 & 0.4 $\pm$  0.2 & 0.22 $\pm$ 0.05 & AM, M \\
J1637+1131 & 16:37:47.6875(14) & +11:31:59.39(4) & 0.5957665944149(10) & $-$9.728(4)E-17 & 53.913(4) & 57485 & 124 & 280 & 1.2 & 4.6 & 56647---58325 & 8.0 & 11.8 & 30.4 & < 25  & < 25 & < 25 & 1.2 $\pm$  0.5 & 0.69  & AM, D, N \\
J1656+0018 & 16:56:31.830(19) & +00:18:04.4(4) & 0.667584921012(10) & $-$9.39(4)E-17 & 47.42(3) & 57685 & 81 & 1742 & 2.8 & 4.3 & 56894---58477 & 8.1 & 11.8 & 30.4 & 2.1 & 3.3 & 4.0 & 0.3 $\pm$  0.2 &  & AM, N, SR \\
J1726$-$0022 & 17:26:27.278(12) & $-$00:22:00.8(3) & 0.764725287739(13) & $-$7.300(8)E-16 & 59.857(18) & 57104 & 68 & 869 & 1.2 & 2.5 & 56647---57562 & 7.2 & 12.1 & 31.3 & 2.2 & 3.5 & 3.3 & 0.4 $\pm$  0.1 &  & \\
J1738+0418 & 17:38:24.575(3) & +04:18:15.25(10) & 0.718536989349(4) & $-$9.5(3)E-18 & 23.530(8) & 57323 & 71 & 339 & 2.6 & 2.0 & 56963---57684 & 9.1 & 11.2 & 29.4 & 1.1 & 1.5 & 1.1 & 0.5 $\pm$  0.1 &  & AM, N \\
J1743+0532 & 17:43:04.452(12) & +05:32:11.34(15) & 0.678589822200(10) & $-$2.157(3)E-15 & 55.422(16) & 57450 & 64 & 622 & 0.8 & 1.3 & 57218---57684 & 6.7 & 12.4 & 31.8 & 2.2 & 3.4 & 3.9 & 1.1 $\pm$  0.3 &  & AM? \\
J1749+1629 & 17:49:10.45(3) & +16:29:43.0(5) & 0.432713602497(14) & $-$5.55(5)E-16 & 59.50(4) & 57471 & 43 & 1694 & 1.8 & 1.3 & 57231---57712 & 7.1 & 12.4 & 31.0 & 3.1 & 4.6 & 7.5 & 0.3 $\pm$  0.2 &  & AM, N \\
J1750+0733 & 17:50:37.15(3) & +07:33:11.7(4) & 0.523899070825(18) & $-$1.876(5)E-15 & 55.69(5) & 57450 & 82 & 2195 & 7.3 & 1.3 & 57218---57684 & 6.7 & 12.6 & 31.6 & 2.3 & 3.4 & 3.8 & 1.3 $\pm$  0.3 &  & AM, N \\
J1802+0344 & 18:02:38.258(5) & +03:44:30.72(11) & 1.505343104273(20) & $-$4.1522(5)E-15 & 76.942(14) & 57000 & 84 & 527 & 1.0 & 2.8 & 56496---57504 & 6.8 & 12.1 & 32.4 & 2.8 & 4.2 & 5.1 & 0.4 $\pm$  0.1 &  & \\
J1807+0359 & 18:07:15.858(3) & +03:59:29.26(5) & 1.251928081695(8) & $-$1.1578(19)E-16 & 52.694(5) & 57029 & 100 & 280 & 1.2 & 2.9 & 56496---57562 & 8.2 & 11.4 & 30.8 & 2.0 & 2.7 & 2.4 & 1.2 $\pm$  0.6 & 0.22  & AM \\
J1832+2749 & 18:32:18.9796(15) & +27:49:36.29(3) & 1.5830130550692(9) & $-$2.81755(7)E-15 & 47.361(5) & 57941 & 98 & 186 & 1.8 & 4.6 & 57110---58772 & 6.9 & 11.9 & 32.2 & 2.7 & 3.4 & 3.9 & 2.5 $\pm$  0.5 &  & AM, D? \\
J1912+1947 & 19:12:51.077(12) & +19:47:40.38(15) & 0.420841476666(7) & $-$2.542(3)E-15 & 94.19(4) & 58049 & 71 & 1079 & 1.7 & 1.2 & 57826---58273 & 6.4 & 12.8 & 31.6 & 4.0 & 6.6 & 3.3 & 0.8 $\pm$  0.3 &  & AM, D?, N \\
J1917+3115 & 19:17:33.045(4) & +31:15:05.63(6) & 0.5434273079951(15) & $-$7.0485(11)E-16 & 81.091(12) & 58329 & 98 & 465 & 3.0 & 2.8 & 57826---58834 & 7.1 & 12.3 & 31.2 & 4.3 & 5.7 & 7.1 & 1.0 $\pm$  0.3 &  & AM, N \\
J1937$-$0023 & 19:37:02.3896(15) & $-$00:23:40.68(7) & 4.163380473766(7) & $-$1.99147(18)E-15 & 67.882(4) & 57634 & 107 & 228 & 2.9 & 6.2 & 56496---58772 & 7.5 & 11.2 & 32.5 & 2.6 & 3.8 & 3.4 & 0.9 $\pm$  0.4 & 0.17 $\pm$ 0.06 & \\
J1938+1505 & 19:38:07.62(7) & +15:05:46.8(10) & 0.34453488326(6) & $-$2.86(8)E-16 & 74.44(14) & 57278 & 62 & 4878 & 2.4 & 1.2 & 57053---57504 & 7.3 & 12.4 & 30.6 & 3.5 & 4.2 & 2.6 & 0.5 $\pm$  0.4 &  & D, N \\
J1942+0147 & 19:42:13.936(9) & +01:47:59.7(2) & 0.711856908341(4) & $-$1.657(4)E-16 & 133.19(3) & 57912 & 109 & 1081 & 2.3 & 4.7 & 57053---58772 & 7.8 & 11.8 & 30.7 & 5.6 & 9.2 & < 25 & 1.7 $\pm$  1.0 & 0.22  & D$^{*}$, N \\
J1945+0720 & 19:45:44.833(3) & +07:20:40.75(7) & 0.931069241340(3) & $-$1.02849(7)E-15 & 62.280(8) & 57634 & 104 & 402 & 7.9 & 6.2 & 56496---58772 & 7.2 & 12.1 & 31.6 & 2.8 & 3.8 & 2.6 & 1.2 $\pm$  0.5 & 0.12 $\pm$ 0.02 & AM? \\
J1946+1447 & 19:46:43.16(3) & +14:47:46.7(3) & 0.43815067512(3) & $-$2.548(4)E-15 & 50.51(4) & 57278 & 58 & 1247 & 1.4 & 1.2 & 57053---57504 & 6.4 & 12.8 & 31.6 & 2.8 & 3.4 & 2.1 & 0.5 $\pm$  0.2 &  & AM, N \\
J1957+0724 & 19:57:03.85(3) & +07:24:24.5(4) & 0.1994415276246(15) & $-$1.02122(13)E-15 & 61.76(5) & 58011 & 73 & 1560 & 0.4 & 4.3 & 57231---58792 & 6.5 & 13.1 & 30.9 & 2.9 & 4.0 & 3.2 & 0.2 $\pm$  0.1 &  & AM \\
J2050+1820 & 20:50:54.04(5) & +18:20:36.3(3) & 0.198085342770(5) & $-$3.8515(14)E-16 & 63.06(3) & 57359 & 66 & 1108 & 0.9 & 4.0 & 56627---58093 & 6.9 & 12.9 & 30.5 & 3.8 & 4.9 & 7.5 & 0.5 $\pm$  0.2 &  & AM, D?, N \\
J2059+1100 & 20:59:04.675(5) & +11:00:28.04(10) & 1.048373857300(5) & $-$2.358(3)E-16 & 60.298(8) & 57520 & 52 & 261 & 0.9 & 3.1 & 56948---58093 & 7.8 & 11.7 & 31.0 & 4.2 & 5.5 & < 25 & 0.7 $\pm$  0.1 &  & D, N \\
J2105+0757 & 21:05:26.193(19) & +07:57:57.9(5) & 0.266917176516(3) & $-$2.6754(10)E-16 & 52.48(5) & 57667 & 86 & 1950 & 1.0 & 4.1 & 56917---58417 & 7.2 & 12.6 & 30.5 & 3.5 & 4.7 & < 25 & 0.6 $\pm$  0.2 &  & D, N \\
J2151+1918 & 21:51:11.459(10) & +19:18:12.54(15) & 0.967338765175(6) & $-$1.6137(8)E-15 & 30.950(18) & 58053 & 54 & 703 & 1.5 & 2.3 & 57628---58478 & 7.0 & 12.1 & 31.8 & 1.9 & 2.4 & 3.0 & 0.2 $\pm$  0.1 &  & AM \\
J2202+2134 & 22:02:16.9853(19) & +21:34:33.88(5) & 0.7362111351584(12) & $-$6.525(8)E-17 & 17.763(8) & 58360 & 114 & 332 & 0.5 & 2.6 & 57888---58834 & 8.2 & 11.6 & 30.3 & 1.3 & 1.4 & 1.4 & 0.5 $\pm$  0.2 &  & AM, N \\
J2252+2455 & 22:52:19.036(4) & +24:55:52.17(8) & 0.5562019391241(19) & $-$3.457(3)E-16 & 34.788(13) & 58506 & 86 & 447 & 2.2 & 1.8 & 58180---58834 & 7.4 & 12.2 & 30.9 & 2.1 & 2.9 & 4.0 & 0.5 $\pm$  0.1 &  & D, N \\
J2329+1657 & 23:29:39.247(11) & +16:57:17.7(3) & 1.582027544328(9) & $-$1.994(3)E-16 & 30.438(20) & 57747 & 107 & 1028 & 2.0 & 5.5 & 56737---58758 & 8.1 & 11.4 & 31.1 & 1.8 & 2.7 & 7.1 & 0.3 $\pm$  0.1 & 0.19 $\pm$ 0.09 & \\
J2340+0831 & 23:40:51.948(3) & +08:31:21.57(10) & 3.297275069476(4) & $-$1.53789(8)E-15 & 23.777(3) & 57412 & 147 & 271 & 3.5 & 5.8 & 56347---58478 & 7.5 & 11.3 & 32.3 & 1.2 & 1.9 & 3.2 & 2.1 $\pm$  1.4 & 0.10 $\pm$ 0.03 & AM, D? \\
J2347+0300 & 23:47:44.471(16) & +03:00:12.1(6) & 0.7214716244018(11) & $-$2.73676(17)E-15 & 16.074(6) & 58179 & 49 & 220 & 1.7 & 3.1 & 57618---58740 & 6.6 & 12.4 & 31.9 & 0.8 & 1.2 & 1.6 & 1.7 $\pm$  0.1 &  & AM, N \\
J2354+0434 & 23:54:08.76(8) & +04:34:04(3) & 1.045575445235(10) & $-$4.785(13)E-16 & 12.58(3) & 58160 & 91 & 1306 & 4.0 & 3.2 & 57580---58740 & 7.5 & 11.8 & 31.3 & 0.6 & 0.9 & 1.1 & 1.5 $\pm$  0.4 &  & AM, N \\

%    \bottomrule
    \end{tabular}
%    \end{adjustbox}
    \caption{AO327 pulsar discoveries with completed timing solutions. Columns are as follows: (1)  the pulsar name based on the timing-derived position, (2--3) the right ascension and declination, (4--5)  the spin and spindown frequencies, (6) the DM,  (7) the reference epoch used for the timing model,  (8--9) the number of TOAs and weighted RMS of the residuals, (10) the chi-squared $\chi^{2}$ of the timing residuals, (11--12) the total time span of the TOAs and the MJD range,  (13---15) the characteristic age $\tau$, inferred surface dipolar magnetic field B, and the spin-down energy loss rate $\dot{E}$, (16--18) distances derived from the NE2001, NE2025, and YMW2016 electron density models (cases of model failure are left empty) (19) average observed 327 MHz flux density and standard deviation, (20) average observed 1380 MHz flux density and standard deviation, and (21) the associated emission features. Numbers in the parentheses are the 1$\sigma$-uncertainties on the last digit as reported by TEMPO2, after weighting the TOAs such that reduced chi-squared $\chi_{red}^{2} = 1$. Flux uncertainties are only reported for cases where more than one detection was made. Emission features by alphabetical ordering are: (D) drifting, ($D^{*}$) bi-drifting, (IP) interpulse, (M) mode-changing, (N) nulling, (AM)  amplitude modulation, (PC) postcursor, (SR) refractive scintillation. In some cases, it remained difficult to make a clear identification of emission features. In such cases, we provide a best guess with a question mark. Raw par files can be obtained at \url{https://github.com/tolszans/AO327/}.}\label{table1}
\end{sidewaystable}

\clearpage

We compare our discoveries with that of GBNCC, a 350-MHz survey which covered nearly the entire sky visible from Green Bank, WV. We note that the timing properties of AO327 discoveries are similar to the 148 timed GBNCC discoveries, as expected given the similar observing frequency and predominantly high Galactic latitude coverage. In particular, we can also estimate the yield, $N_{\rm AO327}$ of AO327 using the final yield of GBNCC, $N_{\rm GBNCC}=148$.

%\internallinenumbers

The total number of discoverable sources, N, approximately scales with the maximum search volume $V_{max}$, which in turn scales with the flux density $S$, and survey minimum detectable flux density $S_{min}$. AO327 was initially planned to cover the entire Arecibo sky excluding the region within $\pm$ 5 degrees of the Galactic plane. Only 65\% was completed, which yields a final solid angle, $\Omega_{\rm AO327} \approx 7600$ deg$^{2}$ while GBNCC sampled the entire GBT sky comprising $\Omega_{\rm GBNCC} \approx$ 32440 deg$^{2}$, giving $\Omega_{\rm AO327}/\Omega_{\rm GBNCC} = 0.23$. With the PUPPI backend, AO327 was approximately three times more sensitive than GBNCC \citep{Deneva_2013}, hence $S_{min,\rm AO327}/S_{min,\rm GBNCC} = 1/3$. To estimate the flux difference at the two frequencies, we assume all pulsars have the population median spectral index of $\alpha=-1.7$ \citep{spectral_index_pap}, which yields $S_{\rm 327MHz}/S_{\rm 350MHz}$ = 1.12. 
Using \citet{Deneva_2013}'s Eq 2. and scaling by $\Omega_{\rm AO327}/\Omega_{\rm GBNCC}$ to account for the difference in solid angles (rather than the maximum solid angle as used there), this yields
\begin{equation}
    N_{\rm AO327} = {N_{\rm GBCCC}}\frac{\Omega_{\rm AO327}}{\Omega_{\rm GBNCC}} \left[   \frac{S_{\rm327MHz}/S_{\rm 350MHz}}{S_{min,\rm AO327}/S_{min,\rm GBNCC}}\right]^{3/2}
\end{equation}
This calculation yields an $N_{\rm AO327}\approx242$. This suggests we will at least double the number of discoveries after completing candidate sifting.

 \subsubsection{PSR~J0916+0658}
 PSR~J0916+0658 is an isolated pulsar with a period of 40.7 ms, and inferred surface magnetic field strength of 3.2 $\times 10^9$ G. Using the timing-constrained position, we checked for and found no coincident GAIA sources \citep{gaia} or supernova associations \citep{green_sn}. Additionally, we find no evidence of accelerations indicating a companion. Given the large characteristic age of this source, and its position in the $P-\dot{P}$ diagram, we conclude that PSR~J0916+0658 is a partially recycled pulsar.
 
 Based on the criterion set forth by \cite{10.1111/j.1365-2966.2010.16970.x}, PSR~J0916+0658 is a disrupted recycled pulsar (DRP). DRPs are believed to originate as the first-born neutron stars in high-mass binary systems, with accretion taking place between the pulsar and companion. Eventually, the companion undergoes a supernova explosion and imparts a kick on the system. When the kick is of sufficient strength and in the right direction to disrupt the binary, the recycled pulsar becomes isolated and therefore a DRP, while  kicks that fail to disrupt the binary form double neutron star (DNS) systems. DRPs are less numerous than expected \citep{Fiore_2023,Swiggum_2023,10.1111/j.1365-2966.2010.16970.x}. The DRPs and DNS populations share the same evolutionary channel, and so their populations each give better constraints on the formation processes. To date, 19 DNS systems are known, while the number of DRPs is around 20 (including PSR~J0916+0658). It is still unclear why there are not more DRPs as evolutionary models suggest DRPs should be several times more abundant than DNS systems. One suggested explanation is that neutron stars in close, interacting binaries receive much smaller natal kicks that other pulsar population models suggest \citep{10.1111/j.1365-2966.2010.16970.x}. It is also possible that their higher space velocities result in higher distances, and hence lower fluxes. Lastly, it is more difficult to unambiguously distinguish DRPs from younger low-B pulsars.
 
\subsection{Emission Features}

\begin{figure}
    \centering
           \includegraphics[page=1,width=1.0\linewidth]{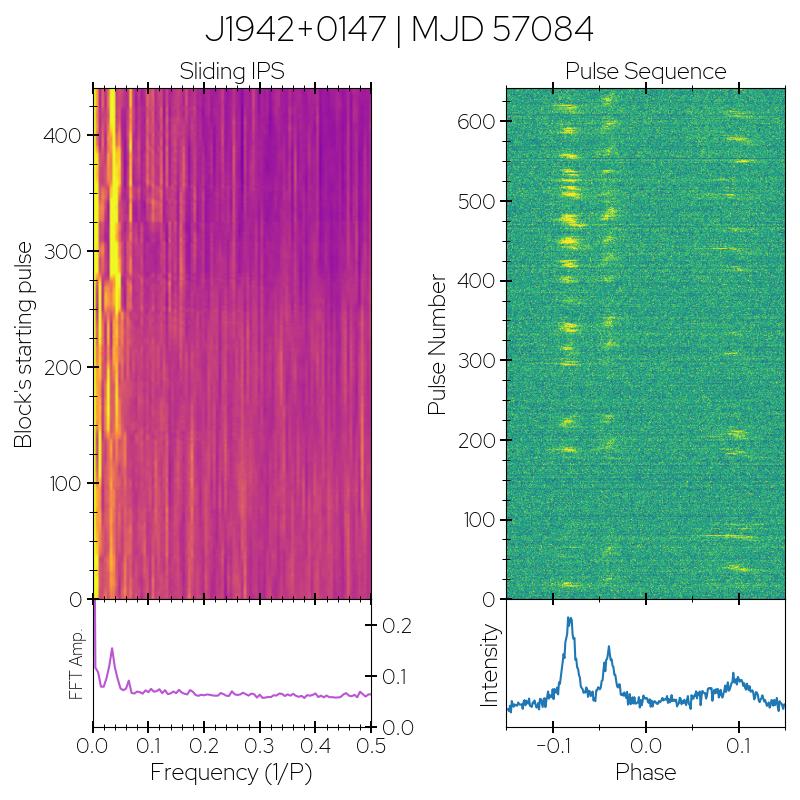} 
     \caption{Top left: Sliding integrated power spectrum for bi-drifting PSR J1942+0147 for our longest observation. Each horizontal slice corresponds to a given data block's (i.e. 200 pulses) integrated power spectrum (as calculated from the LRFS). FFT amplitude is denoted with color. Along the vertical axis we give each data block's starting pulse number. Note the variability of $P_{3}$. Bottom left: The block-averaged integrated power spectrum. Top right: Pulse sequence for the same observation. Note the change in drift direction over the components. Bottom right: The integrated intensity profile. This pulsar has a rather broad profile for its period, comprised of three components, though additional structure is evident in the epoch-averaged profile that could be indicative of a fourth component to the left of the main peak. Four-component profiles are believed to be associated with unusual viewing geometries \citep{1993ApJ...405..285R,cQpaper}.} 
    \label{fig_bd}
\end{figure}

\begin{figure}
    \centering
    \includegraphics[page=1,width=1.\linewidth]{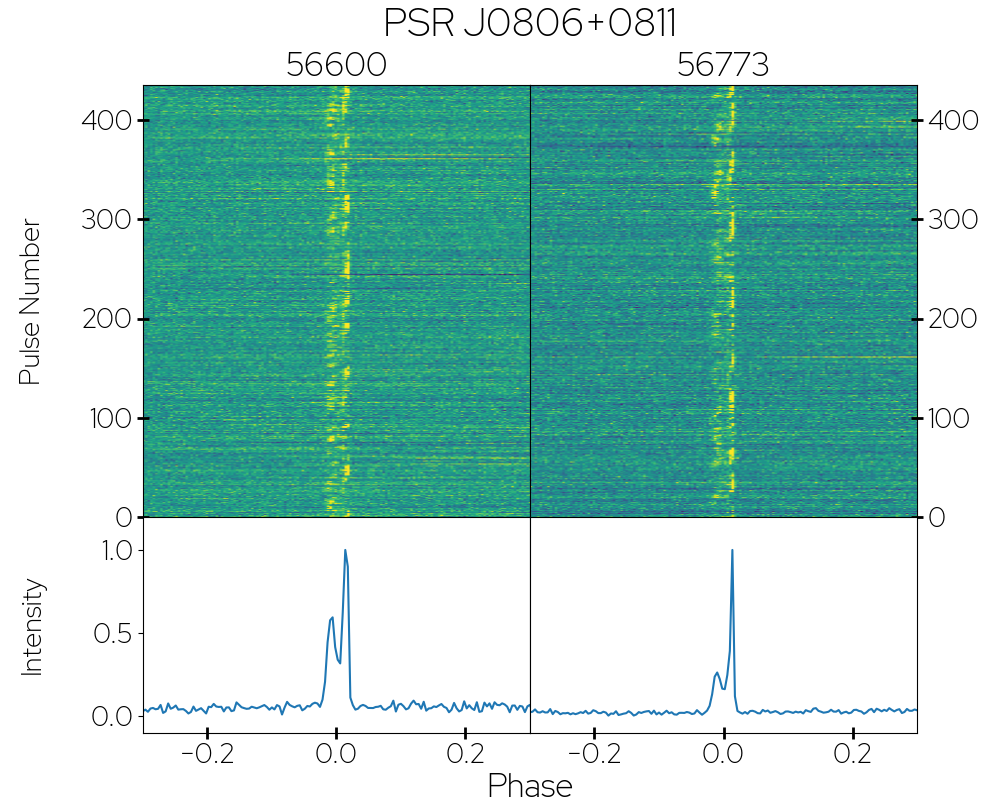}
    \caption{Mode-changing PSR~J0806+0811 observed at 327 MHz. Top: Pulse number vs pulse phase, with color representing intensity. Below: The average profile for this observation. The most significant difference between the two modes can be seen from the relative intensity of the two components. The leading component appearing more often in single pulses on MJD 56600 results in a brighter component.
    }
    \label{fig_mc}
\end{figure}

In addition to constructing phase-connected timing solutions, our multi-epoch observations with single-pulse resolution provide us an opportunity to probe related emission behaviors for these sources. In total, 29 of our sources show some form of amplitude modulation alone, one source shows drifting subpulses alone, and 13 show evidence of both. In addition, PSR J1942+0147 was found to exhibit bi-drifting, a rare type of subpulse drift in which the drift slopes have different signs for different components. This is seen in only a handful of sources \citep{Szary_2020}. An example of this pulsar's bi-drifting is shown in Fig.~\ref{fig_bd}, while an example of mode changing behavior by PSR J0806+0811 is shown in Fig.~\ref{fig_mc}. Documented emission features are listed in Table~\ref{table1}. In cases where identification was uncertain, entries are marked with a question mark.  Polarization calibration was undertaken for a single source, PSR J0225+1727. A detailed study of these pulsars' polarization and  other emission features will be left to forthcoming work. The 327-MHz and L-band epoch-averaged intensity profiles for each pulsar are shown in Fig.~2.1 --~2.6.

\subsubsection{PSR~J0225+1727}
\begin{figure}
    \centering
    \includegraphics[page=1,width=1.\linewidth]{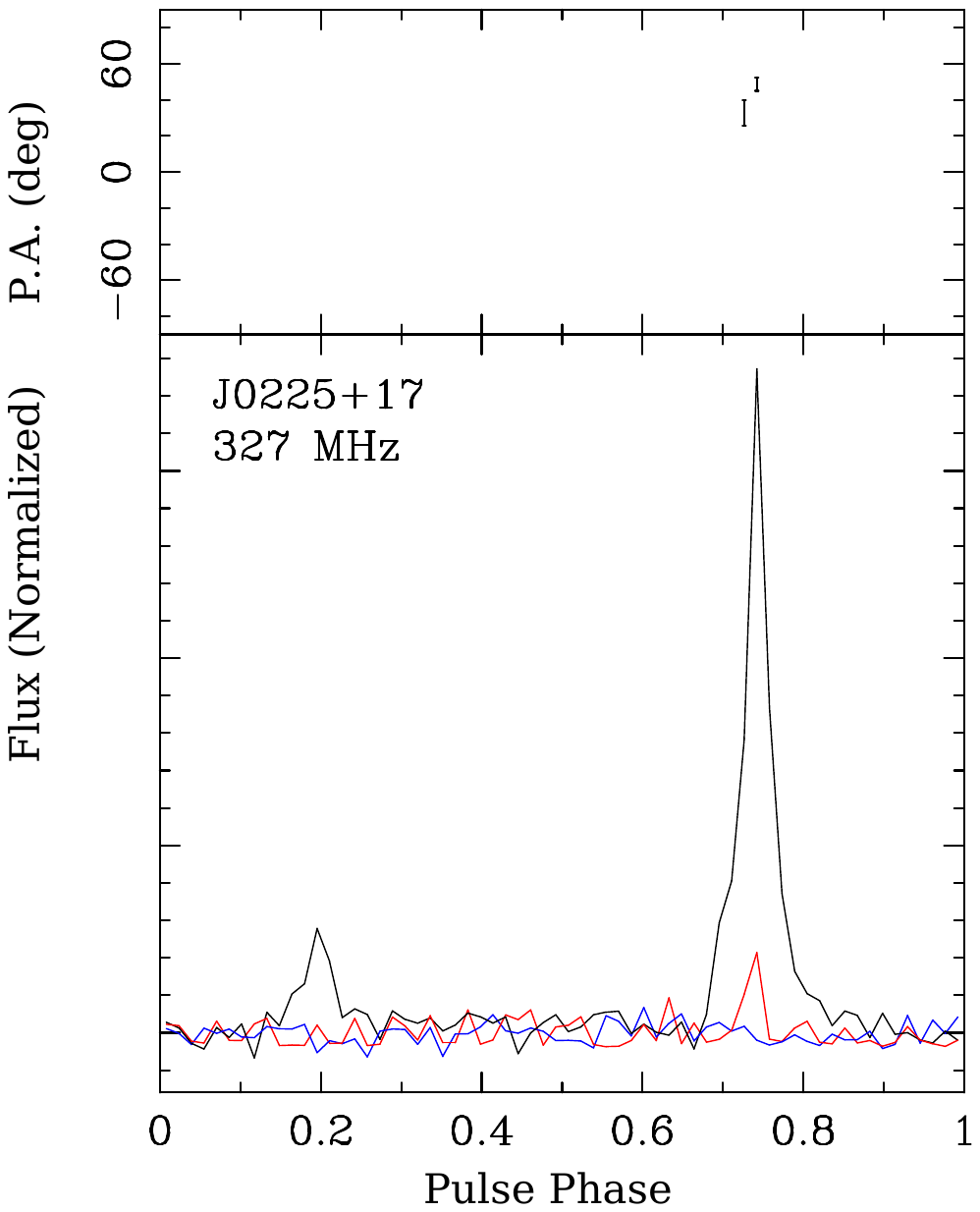}
    \caption{ Top: The average PPA track for PSR J0225+1727. Bottom: Polarization-calibrated composite profile for PSR J0225+1727. The weighted mean RM and weighted mean error are 21.7 $\pm$ 0.3 rad m$^{-2}$. Profiles are normalized, with black lines representing total intensity (Stokes I ), red lines linear polarization (Stokes L), and blue lines circular polarization (Stokes V).}
    \label{pol_plot}
\end{figure}
J0225+1727, with spin period of ~390 ms, was found to have an interpulse offset by 164$^\circ$ from the main pulse. Though the main-pulse of PSR~J0225+1727 was weakly detected at L-band, no detection of the interpulse was made. Because polarization can be used to constrain emission geometry \citep{2001ApJ...553..341E}, it can help determine whether an interpulse originates from the same or opposite magnetic pole as the main pulse. We therefore present a polarization profile for this pulsar in Fig.~\ref{pol_plot}. The weighted mean RM and weighted mean error are 21.7 $\pm$ 0.3 rad m$^{-2}$. After correcting for Faraday rotation, the observations were averaged in time and frequency to yield the composite profile. As noted in Table~\ref{table1}, this source exhibits refractive scintillation. Unfortunately, a significant majority of fold mode observations -- the ones taken in full-Stokes mode -- were taken at epochs where emission was significantly dimmed by scintillation or observations were too short to yield a sufficient detection of the IP necessary for measuring polarization. Because of the low S/N of the resultant profile, and the low polarization fraction in both the main pulse and interpulse, it is unfortunately impossible to place any constraints on the geometry from this profile.

\subsubsection{PSR~J1942+0147}

To investigate the bi-drifting of PSR~J1942+0147, we use an analysis similar to \citet{subpulse_drift} and others \citep{welte2016,2018A&A...616A.119B,10.1093/mnras/staf1710}. We folded the dedispered time series to form the pulse sequence $I(\phi,n)$, where $\phi$ is pulse phase, or longitude, and $n$ the pulse number. Since subpulse drift is a type of periodic modulation, we use fast Fourier transforms (FFTs) to characterize it by its  amplitude power spectrum and phase angle (PA) spectrum. The PA represents the phase (in radians) of the corresponding Fourier sinusoid with frequency $f_{i}$, relative to the beginning of the transformed axis's initial domain. For a pulse sequence, we transform along the pulse number axis, with the resultant 2D amplitude spectrum $A(\phi,f_{i})$ called the longitude-resolved fluctuation spectrum (LRFS) and $\theta(\phi,f_{i})$  as the longitude-resolved phase angle spectrum (LRPAS). 

Subpulse-drift is described by three characteristic quantities; $P_{3}$, the number of pulse periods between drift-bands, $P_{2}$, the longitudinal offset in phase between drift-bands, and lastly the drift-rate $D$ which is defined as the inverse slope of the drift-bands, where $D$ has units of longitude per pulse. Drift properties can change between emission components \citep{subpulse_drift} as well as subsequent drift-bands \citep{McSweeney_2022}, resulting in different values of the drift-rate and of $P_{2}$ and $P_{3}$. The classical and still common determination of $P_{3}$ was pioneered by \citet{backer_paper} and utilizes the LRFS, where the $P_{3}$ appears as a modulation feature with frequency $f_{mod}=1/P_{3}$.  
 For a fixed $P_{3}$, the resulting 1D cut in the LRPAS is referred to as the `subpulse phrase track' \citep{welte2016}. It is most common to fit for $P_2$ using individual drift-bands \citep{2004hpa..book.....L} or using the 2D fluctuation spectra \citep[][2DFS]{Edwards_2002}. Lesser known is Backer's phase analysis \citep{backer_paper} which utilizes the subpulse phase track to directly measure $P_{2}$ through the relative drift-rate across the pulse profile using the PA gradient.  We provide a short derivation in Appendix \ref{drift_appendix} of the relationship between the drift-rate and PA gradient, while for more details consult \citet{don_backer_thesis}. 

The LRFS has both benefits and drawbacks. Since it is sampled for each  pulse period $P$, this yields a Nyquist frequency of $P/2$. In the case $P_{3} < 2 P$, the LRFS would yield an apparent aliased $P_{3}$. In some cases, one can use either the harmonic-resolved fluctuation spectra \citep[][HRFS]{10.1046/j.1365-8711.2001.04079.x} or the 2DFS to resolve the true $P_{3}$ and absolute sense of drift\citep{Edwards_2002}. The chief advantage of the LRFS and subpulse phase track is that they are highly sensitive to longitudinal variability in $P_{3}$ and the drift-rate \citep{subpulse_drift}. While this information is also present in the 2DFS \citep{Edwards_2002}, it is not trivial to extract \citep{welte2016}, and contains nearly the same information \citep{es2002}.

We selected our longest observation, comprising 642 pulses, for investigation. The pulse sequence for this observation is shown in the right panel of Fig.~\ref{fig_bd}. There are a number of drift-bands where the bi-drift is evident, but  there are also times when the bi-drifting is not apparent by eye. Also note the variability in $P_{3}$ throughout the observation. At the beginning there is a long delay between drift bands, before appearing to settle into a regular $P_{3}$ towards the latter half of the observation though still with noticeable variation between drift bands (pulses 400 -- 600).

\begin{figure*}
    \centering
    \includegraphics[page=1,width=1.0\linewidth]{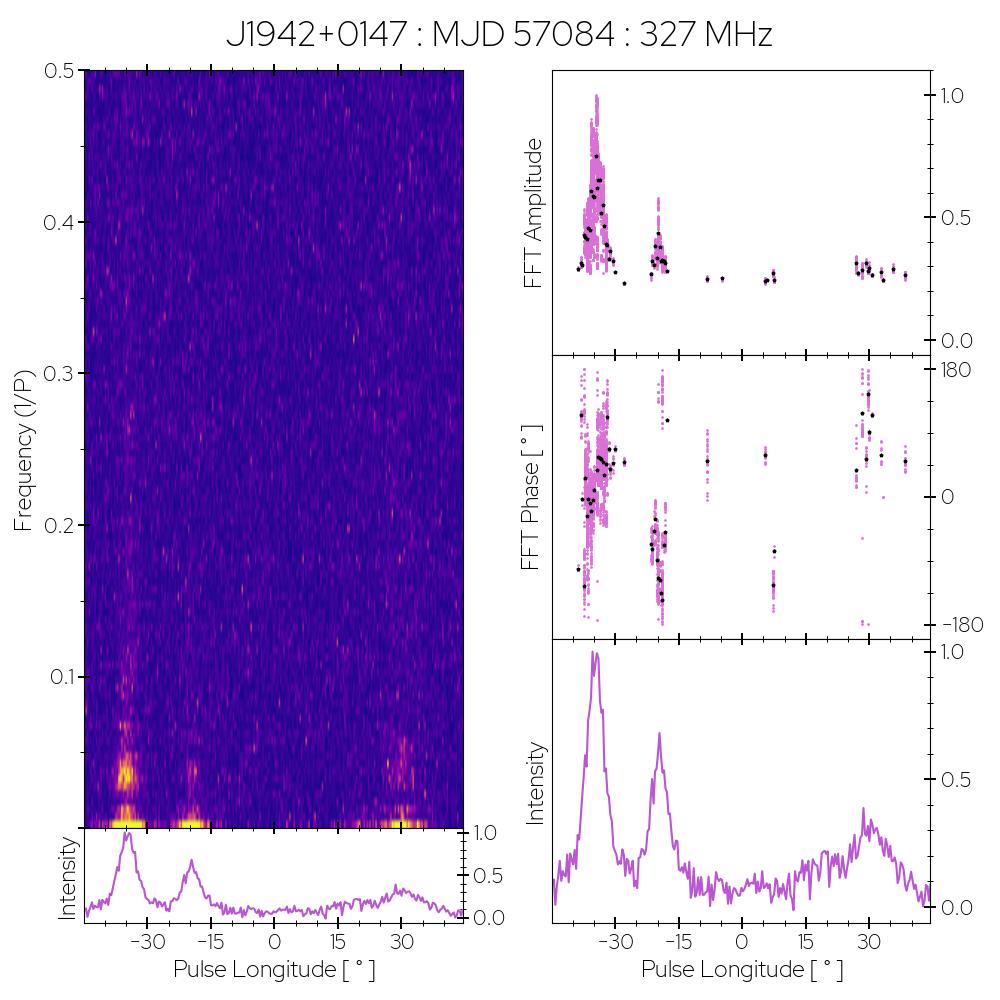}
    \caption{Drift analysis of bi-drifter PSR~J1942+0147 on MJD 57084, our longest observation comprising 642 pulses. Shown to the left is the average LRFS, with the average profile plotted in the panel below. To the right, from top to bottom: modulation amplitude and PA as a function of longitude (purple points denote PA from a given block, while black points delineating the mean), and the average profile. From the average LRFS, it is clear that $P_{3}$ is constant across the profile while the second component exhibits a negative slope in PAs indicating an opposite drift-rate with respect to the other components.}
    \label{bidrift_dip}
\end{figure*}

The pulse sequence was first divided into overlapping contiguous blocks 200 pulses long and offset by intervals of 1 pulse. This is to account for variability in $P_{3}$, which is very common \citep{subpulse_drift}. For each block, a 1D FFT was then applied along the pulse number axis for every longitude bin, forming the LRFS. Across subsequent blocks as well as profile longitude, we normalize each block's LRFS with its maximum amplitude before averaging over all blocks to produce a single block-average LRFS. The block-average LRFS is shown in the left of Fig.~\ref{bidrift_dip}, where we note that $P_{3}=26.3P$ and is stable across the profile. We also note that the $P_{3}$ measured is consistent with the known dependence on $\dot{E}$ \citep{Basu_2025}.

We then longitudinally integrate the block-average LRFS to produce a single average integrated power spectrum (IPS), which we use to estimate the modulation's average peak frequency $f_{avg}$ and its corresponding full width half max (FWHM). The FWHM roughly captures $P_{3}$'s red noise, and in conjunction with $f_{avg}$ can be used to estimate the range of frequencies a given modulation $f_{mod}$ will fall within for a given block and longitude as well as ensuring rejection of secondary modulation features. To highlight the variability of $P_{3}$, for each block's LRFS we calculate the IPS. The `sliding IPS' is shown on the left of Fig.~\ref{fig_bd}. We see that the sliding IPS resolves the $P_{3}$ changes we noted from the pulse sequence.

Next, for each block's LRFS, we search at every longitude bin for a corresponding $f_{mod}$ that falls within the frequency range defined using the average peak frequency and FWHM (i.e. $f_{avg}-5~{\rm FWHM} < f_{mod} < f_{avg}+5~{\rm FWHM}$). To determine whether an $f_{mod}$  value is significant, we  sort the power spectrum from its lowest to highest value, and use the first 15\% of sorted samples to estimate the baseline $b$ while using the remaining 85\% of sorted samples to estimate the baseline RMS, $\sigma_{b}$. If $A(\phi_{i},f_{mod})-b > 5\sigma_{b}$, the peak is considered significant and its amplitude and PA are recorded. 

Between two blocks, the modulation's PAs differ by a fixed offset. To correct for this offset, we first identify from the average LRFS the longitude with the greatest amplitude $\phi_{max}$, and then for each block record the PA associated with $f_{mod}$ at $\phi_{max}$ as the reference PA. For each block's LRFS, the reference PA is subtracted from each measured PA. 

Lastly, the distribution of amplitudes and PAs for all blocks are plotted (shown in the top right and middle of Fig.~\ref{bidrift_dip}) as well as mean amplitudes and PAs. To compute the mean PA, we average together the complex phasor from  each block's $f_{mod}$. We note that the PA gradient (middle right of Fig.~\ref{bidrift_dip}) has a positive slope in components one and four while component two has a negative slope. This indicates component two has an opposite sense of drift as compared to components one and four assuming there is a third component not detectable in single pulses (as is suggested by the epoch-averaged profile). There is still some scatter in the structure of PAs for the leading component, which suggests some deviations from a stable drift-rate. This is similar to what is seen on the right side of Fig.~\ref{fig_bd}, comparing the higher-sloped drift band starting at pulse 490 vs the low sloped band appearing around pulse 515 as well as the change in $P_{3}$ between these two groups of subpulses for the leading component. This suggests the scatter of PAs is caused by a change in $P_{3}$ and $P_{2}$. The other components seem to exhibit a slightly more stable drift-rate, which is consistent with what is seen in Fig.~\ref{fig_bd}.

While single observations suggest a three-component profile, the epoch-averaged profile exhibits additional structure in the trailing component that could be suggestive of a fourth component. We note that this structure corresponds to the longitude range where drift is hardly present. This can be seen in Fig.~\ref{bidrift_dip}; note the interval of 12$\degree$--23$\degree$ which shows weak emission in the average profile but no corresponding significant modulation features. There is a small degree of power in the lowest spectral bin for this longitude range, suggesting some emission exists above noise. A careful investigation of the pulse sequence shown in Fig.~\ref{fig_bd} reveals two pulses (around 90$\degree$ and 450$\degree$) where emission is present for this longitude range. In both cases, the pulses do not appear to be in sync with the drift present at later longitude ranges. Longer observations at 327 MHz and other observing frequencies may lend more clarity to the component structure. It is also worth noting the weak emission that bridges components two and three which is not discernible in single pulses, suggesting this emission is too weak to be resolved  for a single pulse. We note the large profile width, which including the weak wings of emission in the epoch-averaged profile, exceeds 100$\degree$ of phase, which is unusually large given this pulsar's period \citep{tpp_widths}. 

Note that even our longest observation is only 642 pulses. Longer observations, presented in a future work, will  constrain drift properties with more accuracy.

\subsection{Distances and DM Excesses}
\definecolor{MyHexBlue}{HTML}{2EB7CC}
\definecolor{MyHexGreen}{HTML}{2ECC71}
\definecolor{AO327Gold}{HTML}{EFBF04}
\definecolor{GBNCCGold}{HTML}{DAA520}

\begin{figure*}[t]
    \centering
    \includegraphics[page=1,width=.81\linewidth]{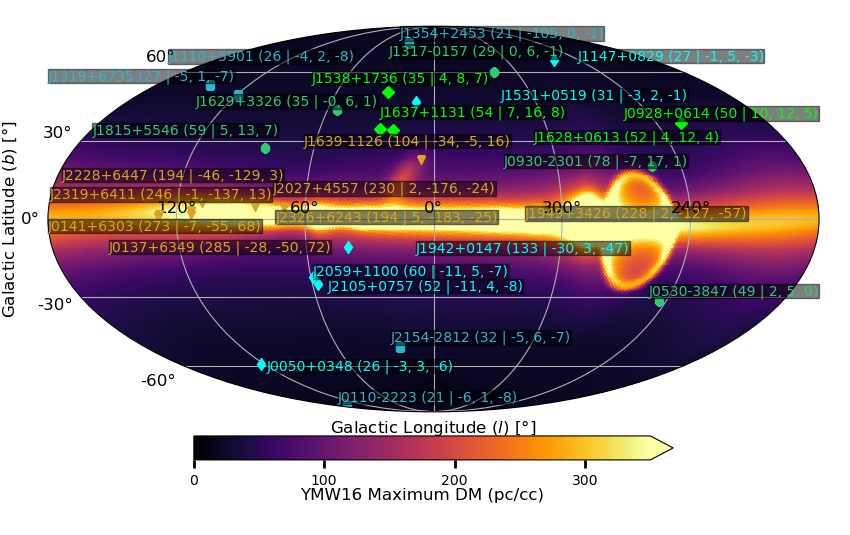}  
     \caption{All-sky map (Mollweide projection) in Galactic coordinates showing a continuous map of YMW16 predicted DMs for sources located at a distance of 30 kpc. Shown are AO327 and GBNCC sources where observed DMs exceed the YMW16 predictions (\textcolor{cyan}{$\blacklozenge$},\textcolor{MyHexBlue}{$\blacksquare$}) where AO327 and GBNCC source DMs exceed YMW16 and either NE2001 or NE2025 (\textcolor{green}{$\medblackdiamond$},\textcolor{MyHexGreen}{$\largeblackcircle$}), and lastly where AO327 and GBNCC source DMs exceed NE2001 or NE2025 failures (\textcolor{AO327Gold}{$\blacktriangle$},\textcolor{GBNCCGold}{$\blacktriangledown$}). Additionally, for each AO327 and GBNCC source we have plotted its name followed in parentheses by source DM, NE2001, YMW16, and lastly NE2025 DM excesses ($\Delta {\rm DM}_{model} = {\rm DM}_{obs} - {\rm DM}_{model}~(d=30~{\rm kpc})$). As all our sources lie on relatively low density lines of sight, for comparison purposes we have limited the color bar to a maximum DM of 350~pc~cm$^{-3}$. In total, 6 AO327 sources see solely YMW16 excesses, while 4 sources see excesses for all three models. In comparison, 10 GBNCC sources see solely YMW16 excesses, 4 sources that exceed all three models, and 8 with NE2001 or NE2025 excesses. We note that a majority of the multi-model excesses are for sources out of the Galactic plane. We also note that the positive latitude YMW16 excesses are also the largest and correspond to cases where both models fail. Interestingly, cases where only NE2001 or NE2025 fail all fall within the Galactic plane.}
    \label{glat_glong}

\end{figure*}

We first explore model predicted maximum DMs for our sources, as shown in Fig.~\ref{glat_glong}. Using the DM excess, defined as $\Delta {\rm DM}_{model} = {\rm DM}_{obs} - {\rm DM}_{model}~(d=30~{\rm kpc})$, we identify 10 sources where the observed DM exceeds the predicted maximum from solely YMW16 (6), or YMW16 and NE2001 and/or NE2025 (4). There are no sources for which only NE2001 and/or NE2025 fail. The sources with model failures are PSRs J0050+0348, J0928+0614, J1147+0829, J1531+0519, J1538+1736, J1628+0613, J1637+1131, J1942+0147, J2059+1100, and J2105+0757. These sources lie predominately off the Galactic plane, and occupy lines of sight not previously sampled. The DM excesses for cases where only YMW16 fails are on average lower than those in cases where two or more models fail. The majority of the DM excesses occur at relatively high latitudes. The largest YMW16 excesses occur for positive latitudes and correspond to cases where NE2001 and/or NE2025 also fail. For sources with DM excesses in negative latitudes, only YMW16 fails. 

While YMW16 was constructed using a larger sample of independent distance measurements (70 through parallaxes, 119 through alternative means) as compared to NE2001 (26 through parallaxes, 86 through alternative means), the uncertainty in non-parallax measurements can be significant. NE2025 has a comparable total number of measurements as YMW16 (171), the uncertainty is far lower due to these being measured through parallax (126) or globular cluster associations (45). Likewise, NE2025 also benefits from a larger pool of DM measurements (4200) over YMW16 (2536) and NE2001 (1143), though even after fitting, $\approx$180 galactic plane pulsars still see excesses\citep{ocker2026ne2025updatedelectrondensity}.
Interestingly, while the DM excesses calculated using NE2025 for our sources are generally smaller than those calculated using NE2021, one source, PSR J0928+0614, shows an increase in the excess.

We repeat this analysis for the entirety of GBNCC's discoveries. We find 10 GBNCC discoveries\footnote{We note that PSR~J1354+2453's published DM is incorrect (private communications). We use the discovery DM, 20 pc cm$^{-3}$, for our analysis.} where the DM exceeds the maximum DM from solely YMW16 (5),  NE2001, or NE2025 (8), or YMW16, and NE2001 and/or NE2025 (3). Similarly to AO327, failures of YMW16 fall in negative latitudes while failures of two or more models fall in positive latitudes. Unlike AO327, GBNCC sources also show cases where only NE2001 or NE2025 fail. The majority of these sources (7) fall in the Galactic plane between $l\approx60\degree-120\degree$ while one falls directly above Galactic Center. Interestingly, three of these are using NE2001 and five are using NE2025, suggesting degraded performance in NE2025 for constraining the maximum density along these lines of sight. For cases where two or more models fail, similarly to AO327, NE2025 shows a smaller DM excess than NE2001's for the majority of sources, though for one source, PSR~J1317+1736, there is an increase in the excess. We also note 8 sources off and on the Galactic plane where NE2025 fails. These are in addition to the $\approx$180 in the Galactic plane identified by \citep{ocker2026ne2025updatedelectrondensity}.

\begin{figure}[t]
    \centering
    \includegraphics[page=1,width=1\linewidth]{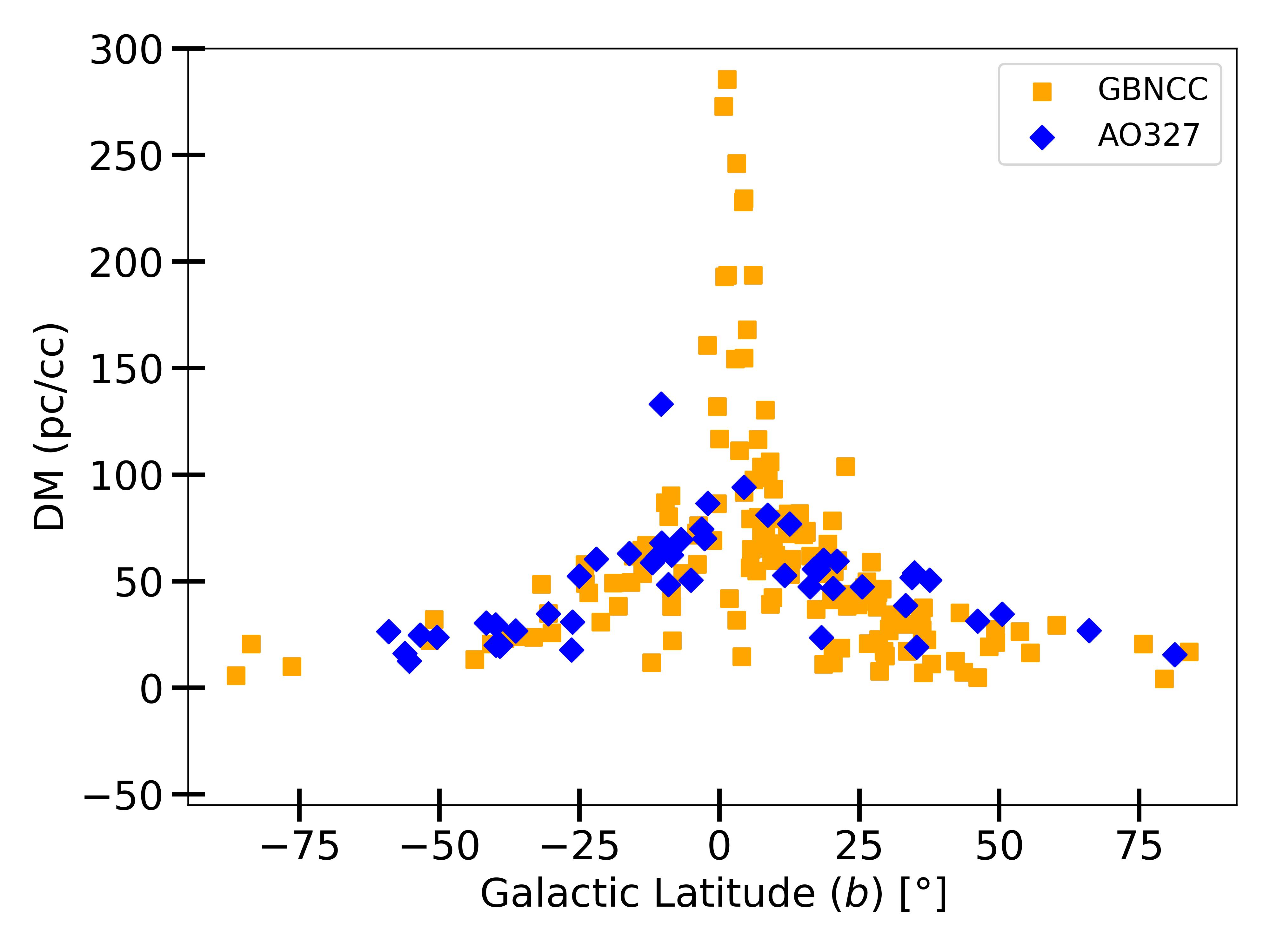}  
    \includegraphics[page=1,width=1\linewidth]{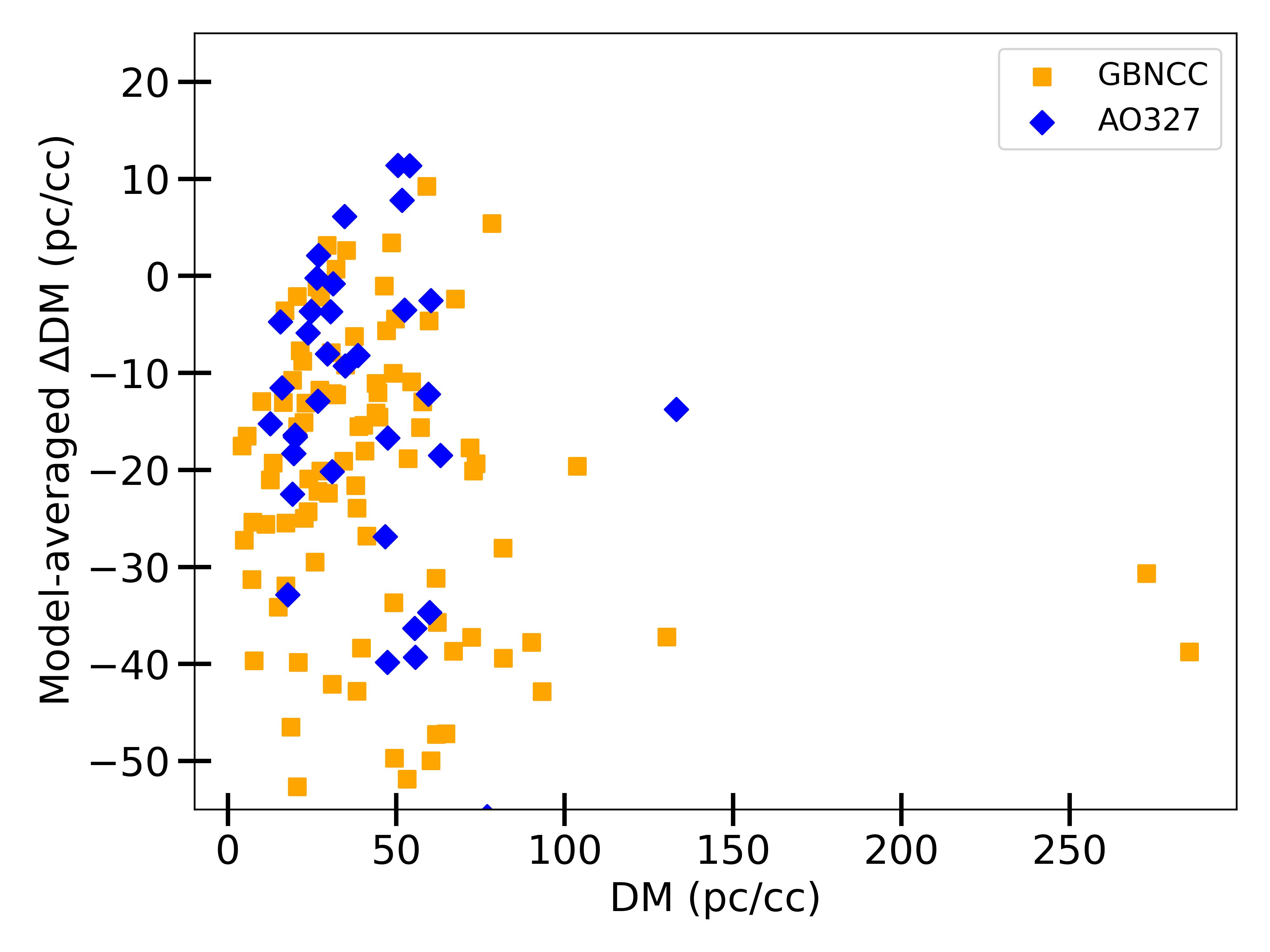}  
    \includegraphics[page=1,width=1\linewidth]{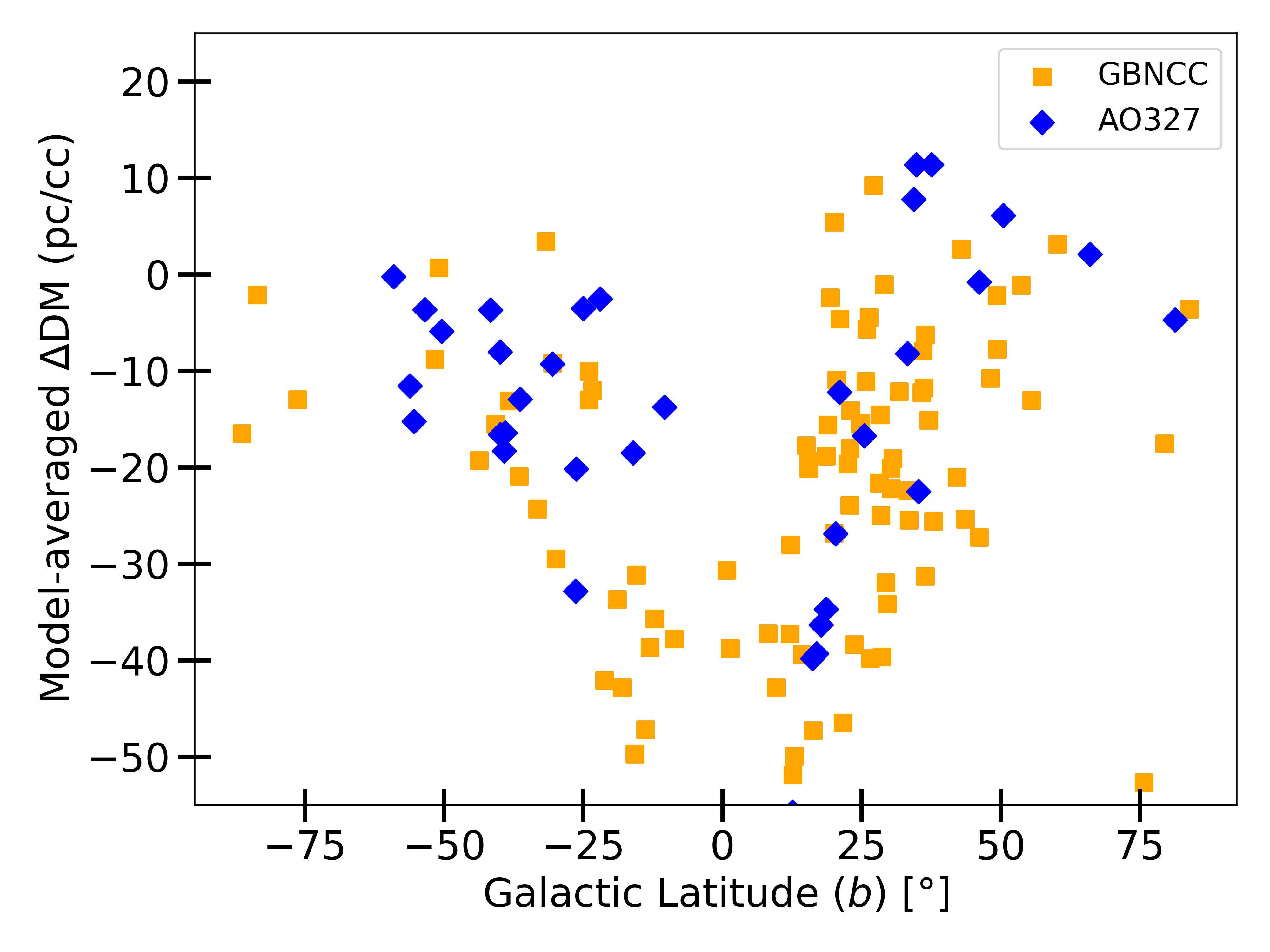}  

     \caption{Top: Pulsar DM vs Galactic latitude for GBNCC (\textcolor{Orange}{$\blacksquare$}), and AO327 (\textcolor{blue}{$\medblackdiamond$}). Middle:  Model-averaged (NE2025 and YMW16) $\Delta $DM vs pulsar DM.  Bottom: Galactic latitude vs model-averaged (NE2025 and YMW16) $\Delta $DM. It is clear that the two surveys exhibit a similar spread of DM and DM excesses.}
    \label{excess}
\end{figure}

To better characterize the relationship with DM, we average  YMW16 and NE2025's DM excess to yield a model-averaged DM excess ($\Delta$DM$_{avg}$). The GBNCC model-averaged DM excess distribution is  mostly in agreement with that of AO327, as shown in the middle of Fig.~\ref{excess}. This is further affirmed by the DM distribution and model-averaged DM excesses as a function of Galactic latitude shown in the top and bottom of Fig.~\ref{excess}. We note that at low Galactic latitudes, GBNCC has more higher DM sources than AO327. This is due to the fact that AO327's sky coverage was mostly concentrated to regions off the Galactic plane ($|b| > 5\degree$)\citep{catalog_paper}.

\begin{figure}[t]
    \centering
    \includegraphics[page=1,width=1\linewidth]{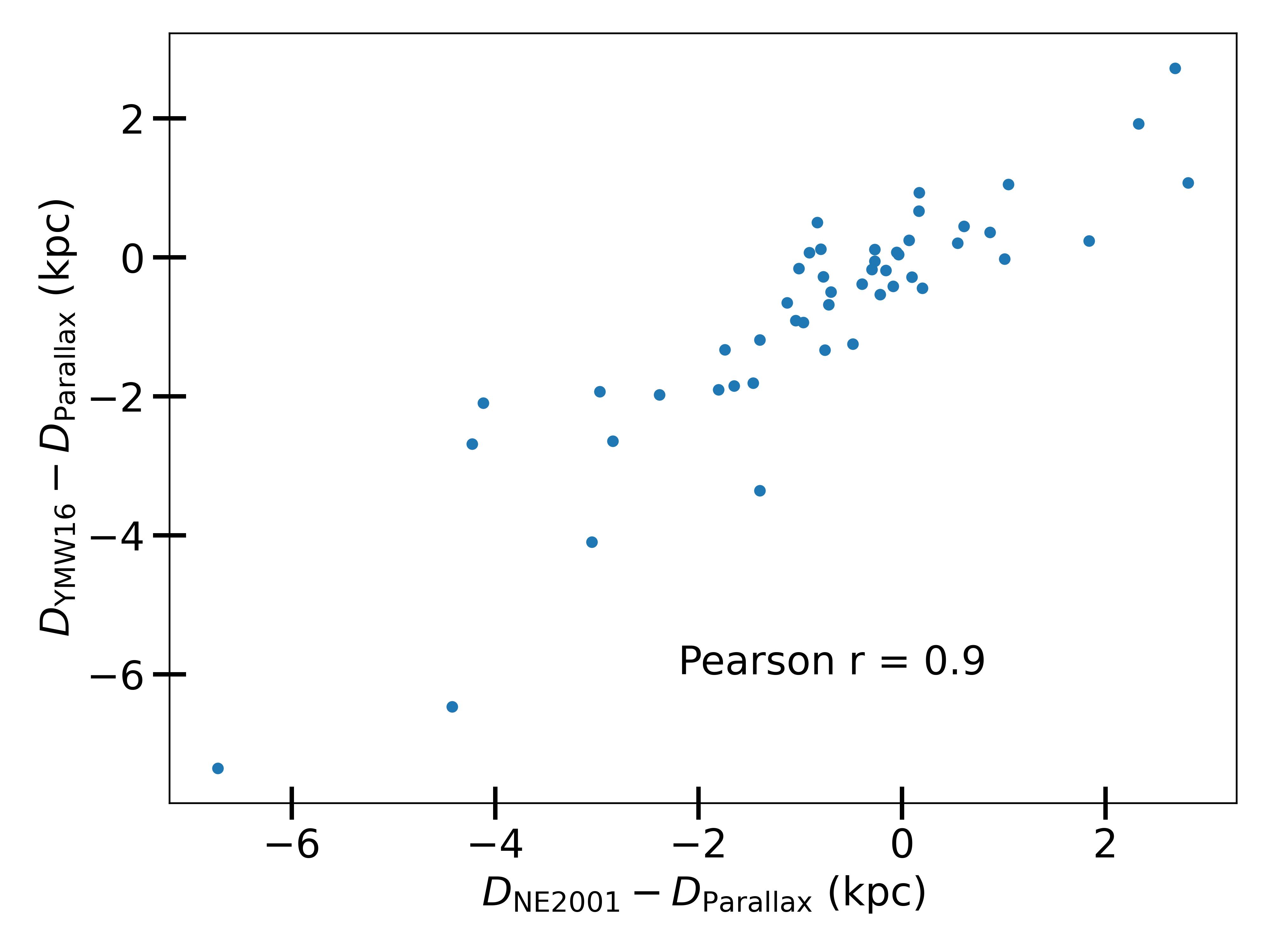} 
    \includegraphics[page=1,width=1\linewidth]{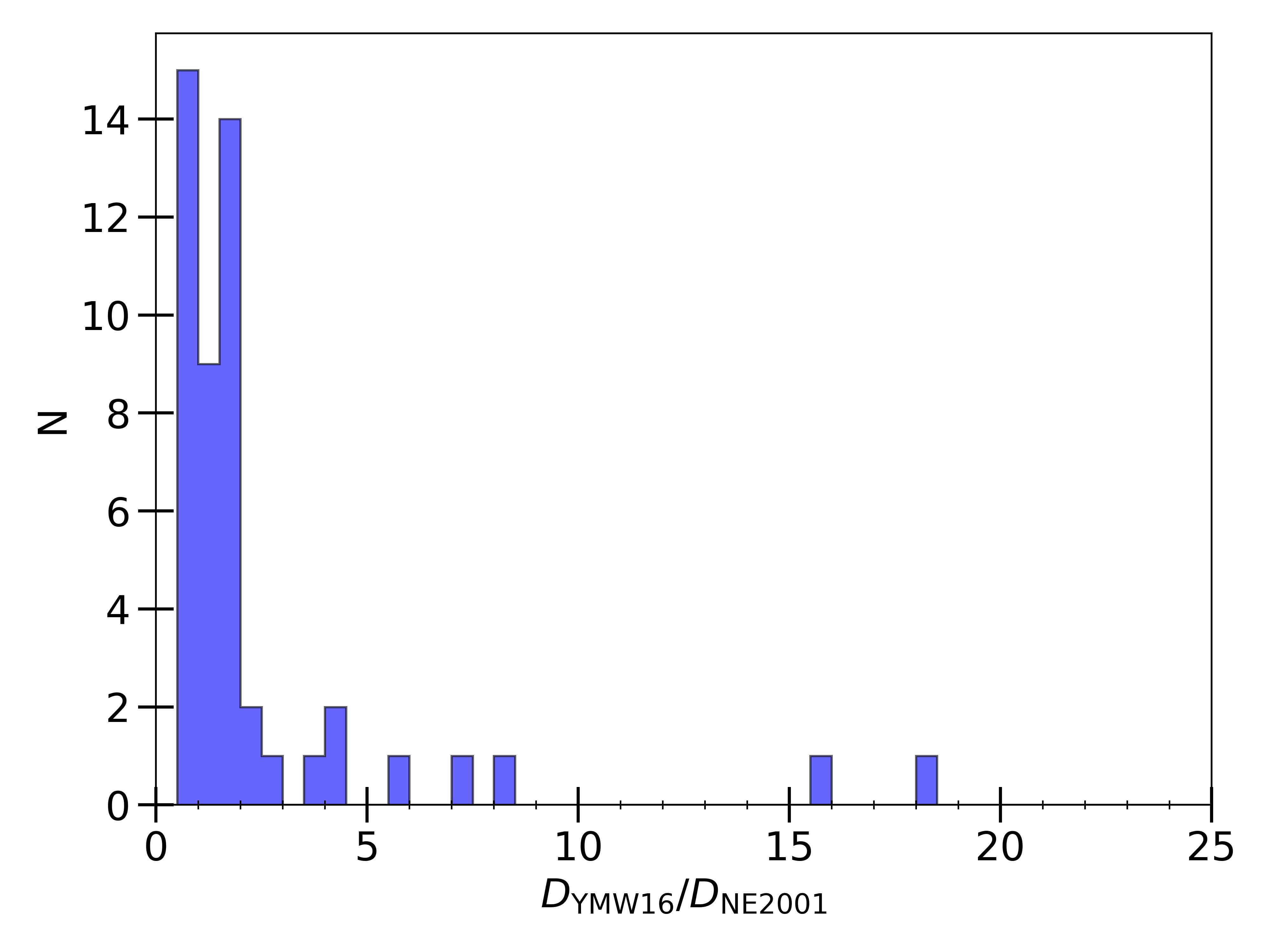} 
    \caption{Top: Scatter plot of model distance uncertainties for sources from \citet{2019ApJ...875..100D} with distance estimates and parallax distances less than 25 kpc. Note the significant correlation between the errors. Bottom: Distribution of $D_{\rm YMW16}/D_{\rm NE2001}$ for this paper's sources. Note the majority of sources cluster around unity, identifying these distance estimates as highly correlated.}
    \label{histo}
\end{figure}

Errors between density models tend to be strongly correlated. This can be directly shown using parallax-derived distances not used in the creation of the models being tested. Shown in Fig.~\ref{histo} are the distance uncertainties between NE2001 and YMW16 for 57 pulsars from the PSR$\pi$ survey with parallax measurements  that were not involved in the creation of either YMW16 or NE2001 \citep{2019ApJ...875..100D}. These pulsars span a wide range of Galactic latitudes and longitudes, and overlap with the AO327 sky. As is clear, the errors are strongly correlated. This has important ramifications for the ratio of the DM-derived distances, $D_{\rm YMW16}/D_{\rm NE2001}$. Since there is significant correlation between distance errors, this will lead to a clustering of sources with the DM-derived distance ratio close to unity as seen in \cite{pygedm}, while the few sources with strongly anti-correlated errors determine the bounds of the distribution. While consistent distance estimates are not inherently useful for inferring true accuracy, inconsistent distance estimates are inherently useful for inferring cases where model performance is problematic. 

Shown in the bottom of Fig.~\ref{histo} is the distribution of the DM-derived distance ratio, $D_{\rm YMW16}/D_{\rm NE2001}$, between YMW16 and NE2001 for AO327 sources. In general, we find that for most of our sources, the distance estimates from the three models are within a factor of two or three of each other. Indeed, the majority of sources cluster near unity with a smaller fraction forming a tail up to large ratios. This tail is made up primarily of previously mentioned sources with DM excesses; PSRs J0050+0348 (18.4), J1147+0829 (15.9), J1531+0519 (8.19), J2105+0757 (7.2), J2059+1100 (5.9), J1942+0147 (4.5), where the numbers in parentheses is the DM-derived distance ratio, $D_{\rm YMW16}/D_{\rm NE2001}$. A few of these sources with DM excesses and highly discrepant distance estimates are close to known pulsars for which model distance estimates agree. For example, PSR~J1942+0147, for which the two distance estimates are very discrepant ($< 25$ kpc using YMW16, 5.6 kpc with NE2001 and 9.2 kpc with NE2025), is only 0.5 degrees away from J1941+0121, for which the distance estimates show excellent agreement (2.1 kpc using YMW16, 2.2 kpc with NE2001, and 3.0 kpc with NE2025). PSR~J1942+0147 has a DM of 133.2 pc~cm$^{-3}$, compared to PSR~J1941+0121's DM of 51.9 pc~cm$^{-3}$, suggesting that the models diverge at large distances, with YMW16, in particular, inaccurate along this line of sight. For PSR J0050+0348, another source with discrepant estimates ($<25$ kpc in YMW16, 1.4 kpc with NE2001, and 2.6 kpc with NE2025), the nearest source PSR J0051+0423 is 0.64 degrees away and shows better agreement between the distance estimates (1.2 kpc using the YMW16 model, 0.6 kpc with NE2001, and 1.0 kpc with NE2025). Again, however, PSR~J0050+0348 has a DM of 26.4 pc cm$^{-3}$ while PSR~J0051+0423's DM is 13.9 pc cm$^{-3}$, similarly suggesting that YWM16 fails at large distances. For other sources, there is no other pulsar within a one degree radius. 

Interestingly, two sources in the bottom of Fig.~\ref{histo} -- PSRs J2329+1657 (4.0), and J0011+0805 (3.9) -- do not possess DM excesses but do exhibit noticeably discrepant distance estimates. Both sources lie significantly below the Galactic plane ($b < -45\degree$). PSR~J0011+0805 has a DM of 24.8 pc~cm$^{-3}$ (with distances of 1.3 kpc using NE2001, 5.0 kpc using YMW16, and 2.0 kpc with NE2025) with the nearest source, PSR~J0023+0923, falling 3.4 degrees away (with a DM of 14.3 pc~cm$^{-3}$,  yielding distances of 0.7 kpc using NE2001, 1.2 kpc using YMW16, and 1.0 kpc with NE2025).  PSR~J2329+1657 has a DM of 30.4 pc~cm$^{-3}$ (1.8 kpc away using NE2001, 7.1 kpc using YMW16, and 2.7 kpc with NE2025), with the nearest source, PSR~J2333+20, falling 3.4 degrees away (with a DM of 12 pc~cm$^{-3}$,  yielding 0.9 kpc using YMW16, 0.8 kpc using NE2001, and 0.9 kpc with NE2025). Interestingly, of three other sources nearby, two (PSRs~J2317+1439 and J2322+2057) show good agreement with distance estimates while one, PSR~J2323+1214, has a YMW16 DM excess. 

It is clear that all three models, to varying degrees, are insufficient at fully modeling the electron density distributions off the Galactic plane. Of the three models, NE2001 and NE2025 yielded the smallest number of DM excesses for GBNCC and AO327 sources while YMW16 DM excesses fell exclusively off-plane. In cases where YMW16 fails, NE2025 is seen to offer smaller DM excesses over NE2001 and thus better constrains the total integrated line of sight density off the Galactic plane. However, we also note several cases off and in the Galactic plane where NE2025 ends up performing worse at constraining the total line of sight density. These differences have important consequences for modeling the intergalactic medium using FRBs.

\section{Discussion} \label{discussion}

Here we discuss some of the individual objects which show interesting emission phenomenology.

\subsection{PSR~J0225+1727}
Pulsar emission is generally confined to a primary window of emission, or a so-called main pulse (MP), with typical duty cycle of a few percent. In rare cases, pulsars show interpulses (IPs), or secondary windows of emission offset from the main pulse by around 180$\degree$ \citep{2004hpa..book.....L}, though this can vary by up to $\sim$30$\degree$. At 327~MHz, PSR~J0225+1727's IP is offset 164$\degree$ from the MP, while no detection is made at L-band. IPs can arise either as a consequence of a double-pole (DP) emission geometry \citep{1977Natur.269..126C}, where the line of sight intersects emission from both magnetic poles, or a one-pole emission geometry, in which both the MP and IP arise from the same magnetic pole due to a nearly aligned rotation and magnetic axis \citep{1977MNRAS.181..761M}. For at least some IPs, fits of the rotating vector model \citep{2001ApJ...553..341E} to polarization position angle (PPA) tracks strongly support a two-pole interpretation \citep{kieth_interpulse}, while in other cases, RVM fits are unable to effectively distinguish between aligned and orthogonal geometries \citep{2001ApJ...553..341E}. In a few of our observations, we note the IP shows infrequent and short in duration($\approx$1 subintegration) bursts of emission .
%Polarization is key to determining the emission geometry.
Unfortunately, due to the low S/N and polarization fraction of the profile, the majority of the PPA track was unresolved. It is possible that longer observations may resolve the full PPA track.

\cite{10.1111/j.1365-2966.2011.18471.x} and \cite{10.1111/j.1365-2966.2010.16970.x} provide the most recent study of commonality of IP emission among the pulsar population. At the time of their writing, the number of known pulsars totaled around 1500 with $\sim$2\% exhibiting IPs as defined as we have. This is roughly consistent with a recent paper on FAST-discovered pulsars, in which 39 of the 682 pulsars have IPs \citep{wang2023fast}. For a purely random distribution of alignments, $\sim$5\% of pulsars should exhibit IPs \citep{10.1111/j.1365-2966.2008.13382.x} as defined as we have, more than twice of what is actually observed. Both \cite{10.1111/j.1365-2966.2011.18471.x} and \cite{10.1111/j.1365-2966.2008.13382.x} were able to reconcile this discrepancy under the condition that the magnetic axis undergoes alignment on $>$10 Myr time-scales, which would also help explain the observed period distribution of pulsars with IPs \citep{10.1111/j.1365-2966.2008.13382.x}. More recently, \cite{beskin} pointed out that counter-alignment was an equally viable explanation as well as predicting period dependencies of both cases for one and two-pole IPs. Determining the emission geometry for PSR J0225+1727 and other new IP pulsars will provide useful constraints for better determining how the magnetic axis evolves with age.
With the recent doubling of IPs from FAST, theorists are well-positioned to improve model constraints.

\subsection{PSR~J1942+0142}
Nearly all pulsars that exhibit subpulse drift have a preferred drift direction over the entire profile. Previously, only three pulsars -- PSRs~J0815+0939, J1034--3224, and B1839--04 -- were known to show the phenomenon of bi-drifting \citep{maura_spd,Szary_2020}, in which the direction of subpulse drift can change between emission components, while only a single pulsar -- PSR~B0826--34 -- was known to exhibit the effect of drift reversal \citep{l+t}. Recently, nine new bi-drifters have been discovered \citep{10.1093/mnras/staf1710}. The characteristics of these new bi-drifters are quite different from PSR~J1942+0142 and past known bi-drifters, showing far narrower profile widths ($<$ 20$\degree$) and smaller $P_{3}$ values. The distribution of $P$, $P_{2}$, and $P_{3}$ among known bi-drifting pulsars currently appears to be consistent with the underlying population of all drifting pulsars.

The cause of subpulse drift is still debated. As described by \cite{l+t}, above the polar cap the electric potential should vary between adjacent field lines, giving rise to a tangential component of the electric field $E_{\perp}$. This then leads to an $\vec{E}\times\vec{B}$ of outflowing plasma, which traditionally\footnote{\cite{l+t} include a review of alternate models to which we refer the interested reader.} has been favored as the cause of subpulse drift. In this picture, variation of electric potential across the polar cap can give rise to bi-drifting \citep{l+t}. \cite{Szary} expanded on this, and noted that in the presence of multipolar fields, the drift no longer centers around the magnetic axis, but instead around the polar cap potential minimum. Interestingly, this also leads to the semi-empirical model of \cite{drift_wright}.

The evidence is suggestive that bi-drift may be more common in the population than previously thought. The nine new bi-drifters studied by \cite{10.1093/mnras/staf1710} are not obvious bi-drifters as apparent from their pulse sequences. It only becomes clear from the PA gradient and $P_{3}$ folding that these pulsars are true bi-drifters. This underscores the importance of PA gradients  in highlighting novel drift behavior including bi-drifting and drift-modes \citep{basu1,10.1093/mnras/staf1710}. In addition to bi-drifters, the switching phase-modulated class of drifters as defined by \citet{subpulse_drift}, are closely related to bi-drifters. Characterized by a change of drift rate between components, in the most extreme cases of this class, one would find all known bi-drifters. It is likely the mechanism responsible for bi-drift is responsible for the entirety of this class's behavior. Investigations of known drifters may uncover more instances of unique drift such as bi-drift or drift-moding and yield further insights into the mechanism that produces drift and bi-drift.

\section{Summary} \label{summary}
We have presented complete timing solutions for 49 pulsar discoveries from the AO327 survey, including one partially recycled pulsar and 46 non-recycled pulsars. They skew towards large characteristic ages and lower DMs as expected for sources found in this type of survey. 

We find 29 pulsars which show some form of amplitude modulation alone, one source showing drift alone, and 13 showing evidence of both. Among these sources, PSR~J1942+0147 exhibits the rare effect of bi-drifting, and J0225+1727 possesses an interpulse. Future studies will explore the emission features in this set of pulsars as well as their polarization properties. 

We evaluate current electron density models with discoveries from the AO327 and GBNCC surveys. Of the three models, the most extensive model failures occur for YMW16, with failures concentrating off the Galactic plane. We also find that NE2025 offers only a marginal improvement off the Galactic plane as compared to NE2001 while showing marginally worse performance in the Galactic plane over NE2001. Lastly, we identify 8 sources on and off the Galactic plane where NE2025 fails.

In total, 105 pulsars have been discovered in the AO327 survey. These are the last of the AO327 discoveries which can be followed up with the Arecibo Observatory.  Currently, all survey search observations have been processed and around 55\% of search candidates remain to be inspected. We expect at least another 100 pulsars to be discovered, and their followup will necessitate other instruments, with the faintest sources requiring the use of the FAST telescope in China. We plan to use advanced RFI mitigation techniques and pulsar search algorithms in future re-processing of the data, which will most likely increase the number of discoveries.

\twocolumngrid

\acknowledgments
We thank the anonymous referee for their thorough and helpful report. M.A.M., E.F.L., and T.E.E.O. are supported by NSF award AST-2009425. J.S.D is supported by NSF award AST-2009335.  M.A.M. is also supported by NSF Physics Frontiers Center award PHYS-2020265. This research was made possible by the NASA West Virginia Space Grant Consortium, Grant \#80NSSC20M0055. The Arecibo Observatory was a facility of the National Science Foundation operated under co-operative agreement by the University of Central Florida and in alliance with Universidad Ana G. Mendez, and Yang Enterprises, Inc. Some of the results in this paper have been derived using the healpy and HEALPix package.

\clearpage
\bibliography{paper}{}
\bibliographystyle{aasjournal}
%\newpage
\appendix

\section{Bi-drifting Analysis}\label{appendix}

\subsection{Relating the Drift-rate to the PA Slope}\label{drift_appendix}
Subpulse drift is described by three characteristic quantities; $P_{3}$, the number of pulses between drift-bands which leads to a modulation feature with frequency $f_{mod}=1/P_{3}$ in the LRFS, $P_{2}$ the longitudinal offset between drift bands, and the drift-rate $D$ which is defined as the inverse slope of the drift bands, 
\begin{equation}\label{a1}
    D = \frac{\Delta \phi}{\Delta t}
\end{equation}
where $D$ has units of longitude per pulse, and $\Delta \phi$ and $\Delta t$ are the longitudinal and pulse-delay corresponding to the subpulse drift. Both $P_{3}$ and $P_{2}$ can show weak variability across longitude depending on the average shape of drift-bands across the emission component, and so for clarity we will define them with longitudinal dependence, i.e $P_{3}(\phi)$, $P_{2}(\phi)$.

We will show that the drift-rate can be related to the PA gradient. To derive the correspondence between the drift rate and the longitudinal variation of $f_{mod}$'s PAs (i.e the subpulse phase track), we make use of the FFT's time-shift property. For a time-series $g(t)$  with $N$ discrete samples, the FFT is equivalent to the decomposition,
\begin{equation}
    g(t) = \sum_{i=1}^{N} A_{i}\cos{(2\pi f_{i}t +\theta_i) }
\end{equation}
where $f_{i}$, $A_{i}$ and $\theta_{i}$ are the frequency, amplitude and PA in radians of the $i$-th Fourier sinusoid. The time-shift property states that shifting a time-series by a delay $\tau$, corresponds to a linear rotation of the unshifted PA spectrum by a frequency-dependent PA $2\pi f \tau$, i.e, 
\begin{equation}
    \mathcal{\hat{F}}\{g(t+\tau)\}(f) = e^{i2\pi f \tau}\mathcal{\hat{F}}\{g(t)\}(f) 
\end{equation}
where $\mathcal{\hat{F}}$ represents the Fourier transform operator. Hence, for a single $i$-th Fourier sinusoid, the relationship between the PA $\theta_{i}$ and time-delay $\tau_{i}$  is
\begin{equation}\label{eq5}
    \theta_{i}=2\pi f_{i} \tau_{i}
\end{equation}

In a pulse sequence, we transform along the pulse number axis. The resultant 2D amplitude spectrum $A(\phi,f_{i})$ is the LRFS, while the 2D phase angle spectrum $\theta(\phi,f_{i})$  is the LRPAS. When subpulse drift is present, the $P_{3}(\phi)$ gives rise to a modulation feature, $f_{mod}(\phi)=1/P_{3}(\phi)$\footnote{Even with weak variability across an emission component, there is little variation between two adjacent longitudes.} 
%which is what is important for this derivation.
across the profile. The corresponding PAs, $\theta(\phi,f_{mod}(\phi))$ comprise the subpulse phase track. The PAs correspond to a pulse-delay via Eq. \ref{eq5}. Hence, a difference in PAs between two neighboring longitude bins ($\phi+\Delta \phi$ and $\phi$) along the subpulse phase track corresponds to the pulse-delay introduced by the subpulse-drift. We can set $\Delta t=\Delta \tau$ for Eq. \ref{a1}, and then use Eq. \ref{eq5} to relate the pulse-delay to the PA gradient $\Delta \theta(\phi,f_{mod})=\theta(\phi+\Delta \phi,f_{mod})-\theta(\phi,f_{mod})$ between two neighboring longitude bins,
  \begin{equation}
      \Delta \theta(\phi,f_{mod}) = 2\pi(f(\phi+\Delta \phi)\tau(\phi+\Delta \phi)-f_{mod}(\phi)\tau(\phi))
  \end{equation}
 This simplifies as $f_{mod}(\phi+\Delta \phi) \approx f_{mod}(\phi)$ giving

   \begin{equation}\label{eq7}
      \Delta \theta(\phi,f_{mod}) = 2\pi f_{mod}(\phi)(\tau(\phi+\Delta \phi)-\tau(\phi))
  \end{equation}
  
  Recalling $f_{mod}=1/P_{3}(\phi)$, and combining Eq. \ref{a1} with Eq. \ref{eq7} results in the final expression from  \citet{backer_paper} for the drift-rate
  \begin{equation}\label{insd2}
      D(\phi) = \frac{2\pi}{P_{3}(\phi)} \left(\frac{\Delta \theta(\phi,f_{mod})}{\Delta \phi}\right)^{-1}
  \end{equation}
where in the limit $\Delta \phi \rightarrow 0$, $\Delta\theta/\Delta \phi \rightarrow d\theta(\phi,f_{mod})/{d\phi}$. This implies, as noted by \citet{p2_measurements},
\begin{equation}
    P_{2}(\phi)=2 \pi  \left(\frac{d \theta(\phi,f_{mod})}{d\phi}\right)^{-1}.
\end{equation}

%\section{Residuals}
%\renewcommand{\thefigure}{B\arabic{figure}}
%\setcounter{figure}{1}
%\input{residuals_2}

\end{document}